\documentclass[aps,prb,reprint,superscriptaddress,amsmath,amssymb]{revtex4-1}
\usepackage{graphicx}
\usepackage{dcolumn}
\usepackage{bm}

\begin{document}


\title{Isotropic parallel antiferromagnetism in the magnetic-field-induced charge-ordered state of SmRu$_4$P$_{12}$ caused by $\bm{p}\,$--$\bm{f}$ hybridization}


\author{T. Matsumura}
\email[]{tmatsu@hiroshima-u.ac.jp}
\affiliation{Department of Quantum Matter, AdSM, Hiroshima University, Higashi-Hiroshima 739-8530, Japan}
\author{S. Michimura}
\affiliation{Department of Physics, Faculty of Science, Saitama University, Saitama 338-8570, Japan}
\author{T. Inami}
\affiliation{Synchrotron Radiation Research Center, National Institutes for Quantum and Radiological Science and Technology, Sayo, Hyogo 679-5148, Japan}
\author{C. H. Lee}
\affiliation{National Institute of Advanced Industrial Science and Technology (AIST), Tsukuba, 305-8568, Japan}
\author{M. Matsuda}
\affiliation{Neutron Scattering Division, Oak Ridge National Laboratory, Oak Ridge, Tennessee 37831, USA}
\author{H. Nakao}
\affiliation{Photon Factory, Institute of Materials Structure Science, High Energy Accelerator Research Organization, Tsukuba, 305-0801, Japan}
\author{M. Mizumaki}
\affiliation{Japan Synchrotron Radiation Research Institute, SPring-8, Sayo, Hyogo 679-5198, Japan}
\author{N. Kawamura}
\affiliation{Japan Synchrotron Radiation Research Institute, SPring-8, Sayo, Hyogo 679-5198, Japan}
\author{M. Tsukagoshi}
\affiliation{Department of Quantum Matter, AdSM, Hiroshima University, Higashi-Hiroshima 739-8530, Japan}
\author{S. Tsutsui}
\affiliation{Japan Synchrotron Radiation Research Institute, SPring-8, Sayo, Hyogo 679-5198, Japan}
\author{H. Sugawara}
\affiliation{Department of Physics, Kobe University, Kobe 657-8501, Japan}
\author{K. Fushiya}
\affiliation{Department of Physics, Tokyo Metropolitan University, Hachioji, Tokyo 192-0397, Japan}
\author{T. D. Matsuda}
\affiliation{Department of Physics, Tokyo Metropolitan University, Hachioji, Tokyo 192-0397, Japan}
\author{R. Higashinaka}
\affiliation{Department of Physics, Tokyo Metropolitan University, Hachioji, Tokyo 192-0397, Japan}
\author{Y. Aoki}
\affiliation{Department of Physics, Tokyo Metropolitan University, Hachioji, Tokyo 192-0397, Japan}


\date{\today}

\begin{abstract}
Nature of the field-induced charge ordered phase (phase II) of SmRu$_4$P$_{12}$ has been investigated by resonant x-ray diffraction (RXD) and polarized neutron diffraction (PND), focusing on the relationship between the atomic displacements and the antiferromagnetic (AFM) moments of Sm. 
From the analysis of the interference between the non-resonant Thomson scattering and the resonant magnetic scattering, combined with the spectral function obtained from x-ray magnetic circular dichroism, it is shown that the AFM moment of Sm prefers to be parallel to the field ($\bm{m}_{\text{AF}} \parallel \bm{H}$), giving rise to large and small moment sites around which the P$_{12}$  and Ru cage contract and expand, respectively. 
This is associated with the formation of the staggered ordering of the $\Gamma_7$-like and $\Gamma_8$-like crystal-field states, providing a strong piece of evidence for the charge order. 
PND was also performed to obtain complementary and unambiguous conclusion. 
In addition, isotropic and continuous nature of the phase II is demonstrated by the field-direction invariance of the interference spectrum in RXD. 
Crucial role of the $p\,$--$f$ hybridization is shown by resonant soft x-ray diffraction at the P $K$-edge ($1s\leftrightarrow 3p$), where we detected a resonance due to the spin polarized $3p$ orbitals reflecting the AFM order of Sm. 
\end{abstract}


\maketitle

\section{Introduction}
Hybridization between localized and itinerant electrons plays an essential role in various kinds of interesting phenomena in $f$  electron systems, ranging from heavy fermion state to orderings of magnetic and higher rank multipole moments.\cite{Portnichenko20} 
The hybridization is in principle dependent on the symmetry relation of the relevant orbitals and therefore has strong relation with these phenomena. 
Importance of understanding the details of orbital dependent hybridization is becoming more important in recent years in studying a wide variety of ordering phenomena including hybrid multipoles.\cite{Hayami18} 
In this paper, we present a rare case where an orbital dependent $p\,$--$f$ hybridization induces a charge order in magnetic fields, which is actually realized in Sm-based filled-skutterudite. 

SmRu$_4$P$_{12}$, a filled-skutterudite compound forming a body-centered-cubic (bcc) lattice of space group $Im\bar{3}$, exhibits an antiferromagnetic (AFM) order with $\bm{q}=(1, 0, 0)$ at $T_{\text N}$=16.5~K,\cite{Lee12,Tsutsui06,Hachitani06a,Ito07,Hachitani06b,Masaki06,Masaki07,Masaki08} accompanied by a metal-insulator transition.\cite{Sekine00,Matsunami05} 
Anomalously, there appears another transition at $T^*$=14~K inside the AFM phase.\cite{Sekine03,Matsuhira02,Matsuhira05,Sekine05,DKikuchi07,Aoki07} 
The anomaly in specific heat at $T^*$ is enhanced with increasing the field and the region of the intermediate phase (phase II, $T^*<T<T_{\text N}$) expands, indicating that the phase II is more stabilized in magnetic fields.  
The origin of this unusual phase has long been a mystery for more than 10 years until a new theory was proposed and verified experimentally.\cite{Shiina13,Shiina14,Matsumura14,Matsumura16}   
The nesting instability for a $\bm{q}=(1, 0, 0)$ charge-density-wave (CDW) of the conduction band, consisting of the $a_u$ molecular orbitals of the $P_{12}$ icosahedra, overcomes the AFM interaction in magnetic fields by incorporating the $4f$ state through the $p\,$--$f$ hybridization, a direct mixing between the $4f$ state of the rare earth and the $p$ state of the surrounding P atoms.\cite{Harima03,Harima08,Shiina10} 
This leads to a field-induced charge order, i.e., a difference in the charge density of the $P_{12}$ molecular orbitals around the Sm ions at the corner (Sm-1) and the center (Sm-2) of the bcc lattice, which is assisted by a staggered ordering of the crystal-field (CF) states of the Sm-$4f$ electrons.\cite{Shiina13,Shiina14,Shiina16,Shiina18} 
To verify this scenario, we have performed resonant x-ray diffraction (RXD) and obtained supporting results for the theory.\cite{Matsumura14} 

The main points of the previous RXD experiment are as follows. 
(1) Atomic displacements are induced in magnetic fields in phase II. 
This has been interpreted as indirectly reflecting the charge order. 
(2) AFM component parallel to the magnetic field is dominant in phase II, giving rise to large and small magnetic moments of the Sm ions. 
This is coupled with the selection of the AFM domains in which the moments are aligned along the $[111]$, $[\bar{1}11]$, $[1\bar{1}1]$, or $[\bar{1}\bar{1}1]$ axis at zero field with a rhombohedral distortion.\cite{Masaki07,Aoki07,Matsumura16}  
In normal AFM orderings, the moments prefer to be perpendicular to the applied field to gain the Zeeman energy. 
However, it is opposite in the phase II of SmRu$_4$P$_{12}$. 
The parallel AFM ($\bm{m}_{\text{AF}} \parallel \bm{H}$) in phase II can be interpreted as a consequence of the staggered ordering of the $\Gamma_7$ and $\Gamma_8$ CF states. 
(3) The directions of the atomic displacements and the magnitudes of the AFM moments are reversed when the field direction is reversed. 
All of these features are consistent with the picture of the field-induced charge order through $p\,$--$f$ hybridization. 

To depict the concept of the field-induced charge order, we drew in Fig.~1 of Ref.~\onlinecite{Matsumura14} a larger magnetic moment on Sm from which the surrounding atoms shift away and a smaller one on Sm to which the surrounding atoms approach. 
However, it is not an experimentally determined picture and is nothing more than a schematic. 
Although the details of the lattice distortion and the atomic displacements were clarified in the subsequent experiment of high precision non-resonant x-ray diffraction, the relation between the magnetic moment of Sm and the local lattice expansion (or contraction) has not been determined yet.\cite{Matsumura16,comment1} 
This information, which is the first goal of the present study, is important to step forward to the fundamental understanding on the mechanism of the charge order through $p\,$--$f$ hybridization. 

For example, in an isostructural PrRu$_4$P$_{12}$ with similar nesting instability with $\bm{q}=(1, 0, 0)$, the charge order occurs at 60 K, below which the nonmagnetic $\Gamma_1$ and magnetic $\Gamma_4^{(2)}$ CF states alternately become the ground state.\cite{Iwasa05a}  
It is experimentally determined that around the Pr site with the $\Gamma_4^{(2)}$ ($\Gamma_1$) ground state with large (small or vanishing) moment the local lattice expands (contracts).\cite{Iwasa05b} 
These pieces of information can be associated with the theoretical analysis that the $\Gamma_1$ orbital has much larger hybridization with the $a_u$ molecular orbital than $\Gamma_4^{(2)}$.\cite{Shiina10} 
Similar alternate ordering of large and small moments, accompanied by a local lattice expansion and contraction, respectively, is also observed in PrFe$_4$P$_{12}$.\cite{Iwasa08,Iwasa12} 

The field reversal method in RXD used in our previous study of Ref.~\onlinecite{Matsumura14} is based on the fact that the intensities for $\pm H$ fields are written as $|F_{\pm}(\omega)|^2=|\pm F_{\text C} + i\alpha(\omega) F_{\text M}|^2$, where $F_{\text C}$ and $F_{\text M}$ represent the crystal structure factor due to atomic displacements (Thomson scattering) and the magnetic structure factor of the AFM order, respectively, for a forbidden reflection with $\bm{q}=(1, 0, 0)$. $\alpha(\omega)$  is a resonance spectral function. 
By measuring the difference in intensity for $\pm H$, we should be able to deduce the relationship between the signs of $F_{\text C}$ and $F_{\text M}$. 
Unfortunately, from the RXD experiment only, however, we could not deduce it because of the unknown phase factor of $\alpha(\omega)$. 

In polarized neutron diffraction (PND), on the other hand, the intensities for up-spin and down-spin incident neutrons are written as $|F_{\pm}|^2=|F_{\text N} \pm r_0 F_{\text M}|^2$, where $F_{\text N}$ is the nuclear structure factor (equivalent to $F_{\text C}$ in RXD) and $r_0=-5.38$~fm is a constant factor. 
In PND, differently from RXD, without an unknown factor $\alpha(\omega)$, we can directly deduce the relationship between the signs of $F_{\text N}$ and $F_{\text M}$, which has been performed for PrRu$_4$P$_{12}$ and PrFe$_4$P$_{12}$.\cite{Iwasa05b,Iwasa12}
We use basically the same method for SmRu$_4$P$_{12}$. 
If we could determine the spectral function $\alpha(\omega)$ in RXD, we should be able to obtain the consistent result from RXD. 
To determine $\alpha(\omega)$ independently, we use x-ray magnetic circular dichroism (XMCD) and x-ray absorption spectroscopy (XAS). 


The second aim of this study is to detect and verify the difference in charge densities on the two P$_{12}$ cages around Sm-1 and Sm-2, and clarify the relationship between the charge density and the magnitude of the Sm moment. 
This will be a genuinely direct evidence for the theory. For this purpose, we use resonant soft x-ray diffraction (RSXD) at the P $K$-edge, with which the $3p$ electronic state of P is directly investigated through the $1s$--$3p$ resonance. 
If the charge density is different between the two cages, we expect different absorption edges for the two P sites, which would give rise to a resonance peak in the structure factor for this forbidden reflection. 

The third aim of this study is to understand the phase II comprehensively by RXD exhibiting clear signals. 
It has been suggested that the phase II has a very isotropic nature with respect to the applied field direction. 
The parallel AFM is expected to be induced for any field directions in accordance with the underlying order of $p$-electron densities and the $\Gamma_7$--$\Gamma_8$ CF levels. This picture should be confirmed experimentally. 


\section{Experiment}
Single crystalline samples were grown by the tin-flux method. The samples used in the RXD and RSXD experiments are the same as those used in Refs.~\onlinecite{Matsumura14,Matsumura16}. 
RXD experiment at the Sm $L_3$-edge was performed at BL22XU in SPring-8. The sample was mounted in a 8~T vertical-field superconducting cryomagnet. 
Polarization of the incident x-ray is in the horizontal scattering plane ($\pi$-polarization).
Polarization analysis of the diffracted x-ray was carried out by using the Cu-220 reflection. 
RSXD experiment at the P $K$-edge, using the same sample, was performed at BL-11B of the Photon Factory in KEK. 
An in-vacuum two-axis diffractometer equipped with a 7~T superconducting magnet\cite{Okamoto14} was utilized without polarization analysis. 
XMCD experiment was performed at BL39XU in SPring-8 by a helicity modulation method using a diamond phase retarder. Powdered sample was prepared by crushing the single crystals. A 7~T cryogen-free superconducting magnet was used to apply magnetic fields parallel to the x-ray beam. 

PND experiment was performed by using the HB-1 triple axis spectrometer (PTAX) at the High Flux Isotope Reactor (HFIR) at Oak Ridge National Laboratory (ORNL). 
The same sample used in Ref.~\onlinecite{Lee12} was used, which is enriched with $^{154}$Sm isotope to avoid severe absorption of neutrons by natural Sm. 
The volume of the sample was less than 1~mm$^3$. 
A vertical magnetic field of 5~T was applied along the $[1\bar{1}0]$ crystal axis, with the $[110]$--$[001]$ horizontal scattering plane. 
A polarized neutron beam of $\lambda=2.46$~\AA\ ($E=13.5$~meV) was selected by a Heusler alloy monochromator. 
The degree of polarization of the incident neutrons was $92.5\pm 1$~\%. 
A Pyrolytic Graphite (PG) analyzer was used to select the energy of the elastically diffracted neutrons and also to reduce the background. 
The sequence of the horizontal collimators was $48^{\prime}$--$80^{\prime}$--$60^{\prime}$--$240^{\prime}$. 
We measured intensities for up ($\parallel \!\bm{H}$) and down ($\parallel \!-\bm{H}$) spin polarizations, which was switched by a Mezei-type spin flipper. 
The actual count rate of the Bragg diffraction from the sample was approximately 1/5 times the previous experiment of Ref.~\onlinecite{Lee12} using unpolarized neutrons from a PG monochromator.

\section{Results and Analysis}
\subsection{AFM domain selection in phase III}
\begin{figure}[t]
\begin{center}
\includegraphics[width=8.5cm]{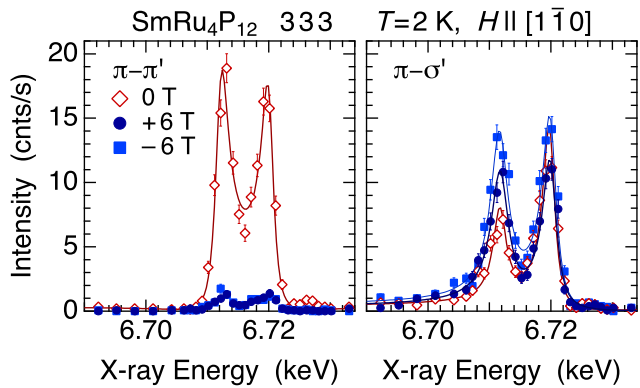}
\end{center}
\caption{X-ray energy dependences of the intensity of the 333 Bragg diffraction at 2 K (phase III) in magnetic fields of 0 and $\pm 6$ T along  $[1 \bar{1} 0]$. After absorption correction.}
\label{fig:Edep2KH110}
\end{figure}
We first describe that the low temperature phase-III ($T<T^*$) is a normal AFM ordered state, where the magnetic moments prefer to be perpendicular to the magnetic field. 
Figure~\ref{fig:Edep2KH110} shows the resonance spectra of the 333 Bragg diffraction at 2 K in magnetic fields of 0 and $\pm 6$ T along  $[1 \bar{1} 0]$. 
Two resonant peaks are clearly observed at $E2$ (6.712 keV, $2p_{3/2} \leftrightarrow 4f$) and $E1$ (6.720 keV, $2p_{3/2} \leftrightarrow 5d$) energies. 
These are of magnetic dipole origin because no signal was observed in the $\sigma$-$\sigma'$ channel at the $E1$ resonance, which was measured by tuning the incident polarization to $\sigma$ by using a phase retarder system.\cite{Inami13,SME1ss} 
The intensity of $\pi$-$\pi'$ scattering decreases significantly by applying the field, whereas that of $\pi$-$\sigma'$ slightly increases. 
This shows the increase in the volume fraction of the AFM domains in which the moments are oriented along the $[111]$ and $[11\bar{1}]$ axes ($\bm{m}_{\text{AF}} \perp \bm{H}$, $\bm{m}_{\text{AF}} \parallel$ the horizontal scattering plane),\cite{Matsumura16} which gives rise to the $\pi$-$\sigma'$ scattering according to Eqs.~(\ref{eq:A-5}) and (\ref{eq:A-7}). 
In contrast, the volume fraction of the $[\bar{1}11]$ and $[1\bar{1}1]$ AFM domains, in which the moments have large parallel component to the field ($\bm{m}_{\text{AF}} \parallel \bm{H}$, $\bm{m}_{\text{AF}} \perp$ the scattering plane) and are responsible for the $\pi$-$\pi'$ scattering, decreases by applying the filed because they are energetically unfavorable. This is quite a normal behavior of an AFM order.  
Almost the same result is obtained also for $H \parallel [\bar{1}\bar{1}2]$.\cite{SM1}

\subsection{Field-induced charge order in phase II}
\begin{figure}[t]
\begin{center}
\includegraphics[width=8cm]{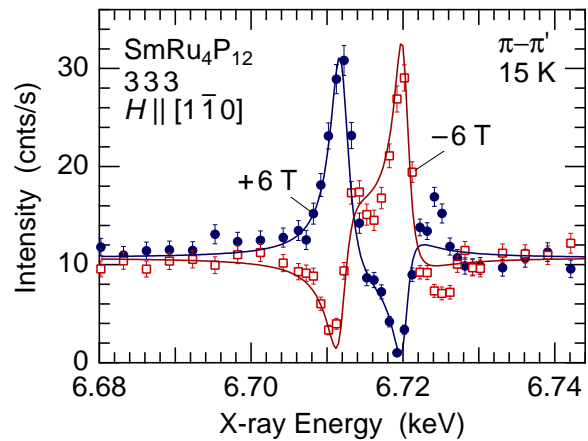}
\end{center}
\caption{(a) X-ray energy dependences of the intensity of the 333 Bragg diffraction for the $\pi$-$\pi'$ channel at 15 K in magnetic fields of $\pm 6$ T along $[1 \bar{1} 0]$. Solid lines are the fits with Eq.~(\ref{eq:IntRXD}), using the magnetic spectral function $\alpha^{(1)}(\omega)$ obtained from XMCD. See text.  
}
\label{fig:Escan15KppH110}
\end{figure}
Figure~\ref{fig:Escan15KppH110} shows the x-ray energy dependence of the 333 Bragg-diffraction intensity for the $\pi$-$\pi'$ channel at 15 K in phase II in magnetic fields of $\pm 6$ T for $\bm{H}\parallel [1 \bar{1} 0]$. 
Although this result has already been reported in Ref.~\onlinecite{Matsumura14} for the 030 reflection in $\bm{H} \parallel [001]$, the present result for the 333 reflection in $\bm{H}\parallel [1 \bar{1} 0]$ is more clear-cut. 
Non-resonant Thomson scattering appears in phase II due to the field-induced atomic displacements, which is considered to reflect the underlying charge order. 
Simultaneously, the resonant magnetic scattering in the $\pi$-$\pi'$ channel is enhanced, indicating the development of the parallel AFM component, which is usually not preferred and is suppressed in phase III. 
The non-resonant magnetic scattering is very weak (less than 1 cps) as can be observed in the nonresonant region below 6.70 keV in Fig.~\ref{fig:Edep2KH110}, which can be reasonably neglected. 
The non-resonant Thomson scattering and the resonant magnetic scattering interfere with each other and exhibit characteristic anomalies around the $E2$ and $E1$ resonance energies. 
The interference structure in Fig.~\ref{fig:Escan15KppH110}, which is reversed by changing the field direction, contains information on the relationship between the atomic displacement and the magnitude of the parallel AFM component of Sm. 
The energy spectrum should be analyzed by using the following function. 
\begin{align}
I(\omega) &= \bigl| F_{\text{C}} + i \{ \alpha_{E2}^{(1)}(\omega) \bm{G}_{E2}^{(1)} + \alpha_{  E1}^{(1)}(\omega) \bm{G}_{E1}^{(1)} \} \cdot \bm{F}_{\text{M}} \bigr|^2\,, \label{eq:IntRXD} \end{align}
\begin{align}
\alpha_{E2}^{(1)}(\omega) &=  \frac{I_2 \Gamma_2 e^{i\phi_2}}{\hbar\omega - \Delta_2 + i\Gamma_2} \,, \label{eq:alpE2} \\
\alpha_{E1}^{(1)}(\omega) &=  \frac{I_1 \Gamma_1 e^{i\phi_1}}{\hbar\omega - \Delta_1 + i\Gamma_1} \,, \label{eq:alpE1} 
\end{align}
where $\alpha_{E2}^{(1)}(\omega)$ and $\alpha_{E1}^{(1)}(\omega)$ are the $E2$ and $E1$ resonance spectral functions due to magnetic dipole moments (rank-1), respectively. $\bm{G}_{E2}^{(1)}$ and $\bm{G}_{E1}^{(1)}$ are the geometrical factors described in the Appendix, which are calculated to be $\bm{G}_{E2}^{(1)}=(-0.77, 0.77, 0)$ and $\bm{G}_{E1}^{(1)}=(-0.68, 0.68, 0)$ in the present geometry of the 333 reflection for $\pi$-$\pi'$ with the $[1 \bar{1} 0]$-axis oriented upwards. 
Although the $E2$ resonance could have a magnetic octupolar (rank-3) contribution, it is estimated to be weak and may be neglected. 
This is justified by the polarization dependence of the $E2$ intensity, which can be explained by the magnetic dipolar geometrical factor.\cite{SM1} 
$F_{\text{C}}$ is the crystal structure factor due to the atomic displacements of P and Ru atoms. The Sm atoms remain on the same sites without breaking the site symmetry and do not contribute to $F_{\text{C}}$ nor other resonances such as the $E1$-$E2$ mixed processes. 
$\bm{F}_{\text{M}}=\bm{m}_1 - \bm{m}_2$ is the magnetic structure factor of the AFM order.   
Only the component of $\bm{F}_{\text{M}}$ along $[1 \bar{1} 0]$ contributes to the $\pi$-$\pi'$ intensity.

\begin{figure}[t]
\begin{center}
\includegraphics[width=8cm]{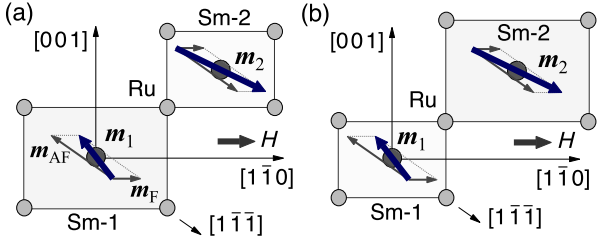}
\end{center}
\caption{Two possible models of atomic displacements in the $[\bar{1} 1 1]$ AFM domain for the field applied along $[1 \bar{1} 0]$. 
We call the Sm atom with its moment oriented to $[\bar{1} 1 1]$ at zero field as Sm-1 and the other one as Sm-2. 
In a magnetic field, a ferromagnetic component $\bm{m}_{\text{F}}$ is induced, resulting in a smaller moment for Sm-1 and a larger moment for Sm-2 ($m_1 < m_2$). 
In (a), the P$_{12}$ cage and the Ru cube expand around the small moment site of Sm-1, whereas in (b) they expand around the large moment site of Sm-2. Only the Ru atoms are shown. 
}
\label{fig:MagstModel}
\end{figure}

Let us consider the atomic displacements in the $[\bar{1} 1 1]$ AFM domain, which is preferred in a magnetic field applied along the $[1 \bar{1} 0]$ direction. 
The two possible cases are shown in Fig.~\ref{fig:MagstModel}. 
Our goal is to determine which case is actually realized. 
However, it is not possible without the knowledge of the phase factors of $\alpha(\omega)$ as shown below. 
If we use the atomic displacement parameters determined in Ref.~\onlinecite{Matsumura16} ($\delta=1.3\times 10^{-4}$ for Ru, $\delta_u=-0.6\times 10^{-4}$ and $\delta_v=1.5\times 10^{-4}$ for P), we have negative $F_{\text{C}}$ for the case (a) ($F_{\text{C}}=-0.331 + 0.0232i$ to be exact) and 
the sign reverses for the case (b). 
By neglecting the small imaginary part of $F_{\text{C}}$, the data in Fig.~\ref{fig:Escan15KppH110} can be fit by tuning $F_{\text{C}}=-3.27$, $I_{2}\bm{G}_{E2}^{(1)}\cdot\bm{F}_{\text{M}}=2.73$, $I_{1}\bm{G}_{E1}^{(1)}\cdot\bm{F}_{\text{M}}=2.52$, $\phi_2=-2.62$ rad, and $\phi_1=0.92$ rad. Other parameters are $\Delta_2=6.712$ keV, $\Delta_1=6.720$ keV, $\Gamma_2=1.5$ eV, and $\Gamma_1=1.3$ eV. 
The calculated curves are shown in Fig.~\ref{fig:Escan15KppH110} by the solid lines.  
The negative $F_{\text{C}}$ here means that the P$_{12}$ cage and the Ru cube expand around the small moment site of Sm-1, i.e., the case (a) is realized. 
It is noted, however, the same fitting curves are obtained by using $F_{\text{C}}=3.27$, $\phi_2=-2.62+\pi$, and $\phi_1=0.92+\pi$, where both of the signs of $F_{\text{C}}$ and $\alpha(\omega)$ are reversed. This set of parameters gives the opposite conclusion for the case (b). 
This is the reason we need to determine the phase of the spectral function by another method. 
The data analysis will be performed again after determining $\alpha(\omega)$ by XMCD.

\begin{figure}[t]
\begin{center}
\includegraphics[width=8cm]{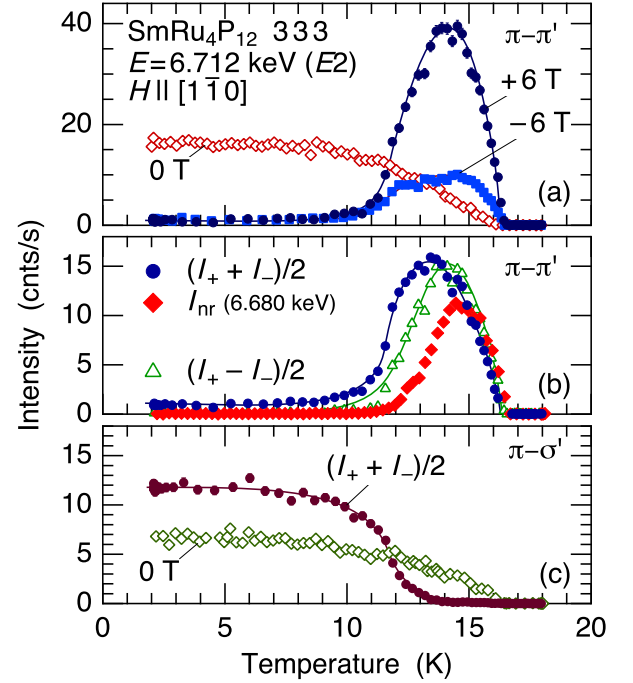}
\end{center}
\caption{(a) Temperature dependences of the 333 Bragg-diffraction intensity for $\pi$-$\pi'$ at $E=6.712$~keV ($E2$) in magnetic fields of 0 and $\pm 6$~T. 
(b) Temperature dependences of the average and difference intensities for the data at $\pm 6$~T and the non-resonant intensity at 6.680~keV measured in the $\pi$-$\pi'$ channel. 
(c) Temperature dependences of the intensity at 0 T and the averaged intensity at $\pm 6$~T for $\pi$-$\sigma'$.  
 The lines are guides for the eye. }
\label{fig:TdepIntppH110}
\end{figure}
\begin{figure}[t]
\begin{center}
\includegraphics[width=8cm]{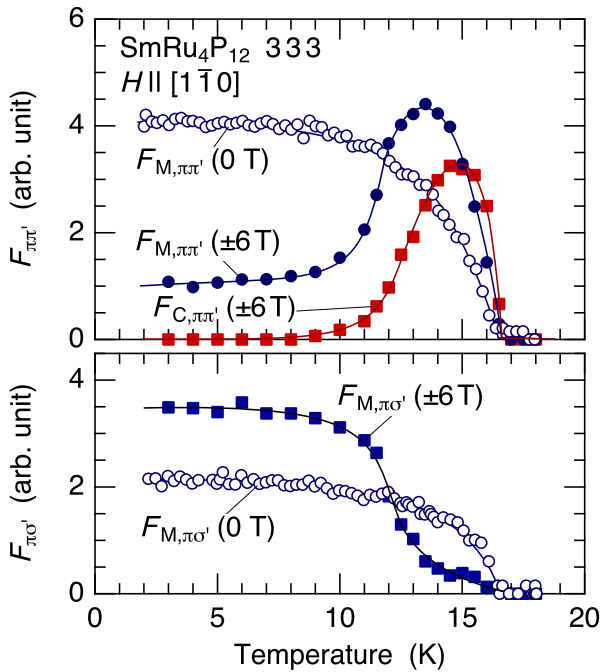}
\end{center}
\caption{Temperature dependences of the crystal and magnetic structure factors for $\pi$-$\pi'$ (top) and $\pi$-$\sigma'$ (bottom), which are deduced from the intensity in Fig.~\ref{fig:TdepIntppH110}. The solid lines are guides for the eye. }
\label{fig:TdepFpsFpp110}
\end{figure}
The $T$-dependences of the $\pi$-$\pi'$ and $\pi$-$\sigma'$ intensities at the $E2$ resonance energy at zero field and $\pm 6$~T are shown in Fig.~\ref{fig:TdepIntppH110}. 
At zero field, both $\pi$-$\pi'$ and $\pi$-$\sigma'$ intensities  exhibit a normal $T$-dependence of an AFM order. 
In magnetic fields, the $\pi$-$\pi'$ scattering exhibits a strong asymmetry with the field reversal in phase II. 
The $T$-dependence of the intensity of the non-resonant term is shown as $I_{\text{nr}}$ in Fig.~\ref{fig:TdepIntppH110}(b), which has been measured at 6.680~keV and 6~T. This is proportional to $|F_{\text{C},\pi\pi'}|^2$. 
The development of the parallel AFM component on entering phase II is clearly detected as the enhancement of the average intensity $(I_{+} + I_{-})/2$, reflecting $|F_{\text{M},\pi\pi'}|^2$. 
The difference in intensity $(I_{+} - I_{-})/2$ shows the $T$-dependent interference effect, which is proportional to $F_{\text{C},\pi\pi'}F_{\text{M},\pi\pi'}$. 
In the $\pi$-$\sigma'$ channel, on the other hand, there arise little asymmetry in the field reversal and only the averaged intensity is shown in Fig.~\ref{fig:TdepIntppH110}(c). 
It is noted, however, the $\pi$-$\sigma'$ intensity, reflecting the perpendicular AFM component, is suppressed in phase II. 
This result consistently shows that the parallel AFM component is enhanced in phase II. 
Although these features had been predicted from our first report for $\bm{H}\parallel [001]$, the present results demonstrate the anomalous magnetic state in phase II more clearly. 

The $T$-dependences of the crystal and magnetic structure factors deduced from the intensity data are shown in Fig.~\ref{fig:TdepFpsFpp110}, which directly demonstrates the behavior of the order parameters for $\bm{H} \parallel [1 \bar{1} 0]$. 
The normal $T$-dependence of the AFM order parameter at zero field changes by applying a magnetic field, forming a field-induced phase below $T_{\text{N}}$. 
The parallel AFM component develops as represented by the enhancement of $F_{\text{M},\pi\pi'}$ and by the suppression of $F_{\text{M},\pi\sigma'}$, accompanied by a staggered atomic displacement represented by $F_{\text{C},\pi\pi'}$. 
With further decreasing temperature, these features disappear and the normal AFM phase is recovered, where the perpendicular AFM domains dominate. 

\subsection{Isotropic ordered state in phase II}
\begin{figure}[t]
\begin{center}
\includegraphics[width=8.5cm]{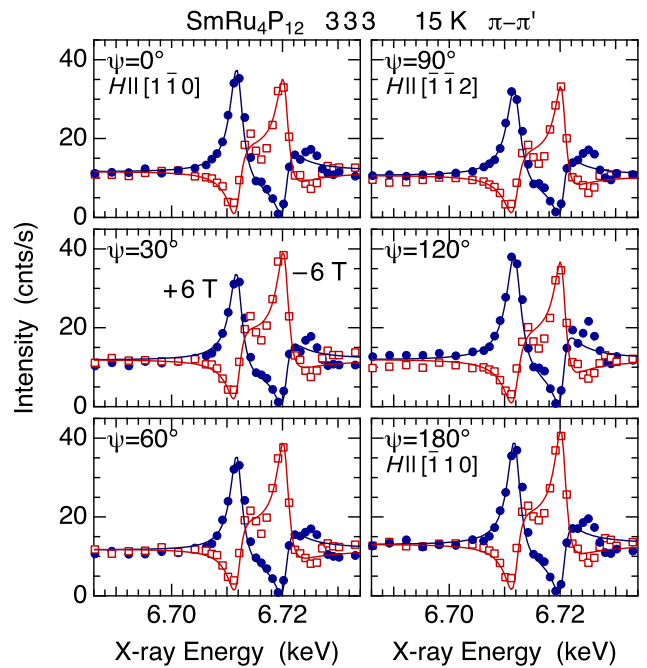}
\end{center}
\caption{Energy spectrum of the 333 Bragg-diffraction at $\pm 6$~T in phase II measured as a function of the field direction, as represented by the azimuthal angle $\psi$. }
\label{fig:Azi6T333}
\end{figure}

One of the distinctive features of this field-induced phase is its continuous response of the order parameter to the magnetic field, resulting in an isotropic nature. 
Figure~\ref{fig:Azi6T333} shows the field-direction (azimuthal angle) dependence of the energy spectrum measured by rotating the sample about the scattering vector (3,~3,~3) in magnetic fields of $\pm 6$~T in phase II at 15~K. 
The magnetic field was first set at $+6$~T and the energy scans were carried out with the rotation of the sample, i.e., the field direction was changed by a step of 30$^{\circ}$ in a constant field of $+6$~T. 
As clearly demonstrated in Fig.~\ref{fig:Azi6T333}, the energy spectrum does not change with the rotation. 
Although $+6$~T at $\psi=180^{\circ}$ is geometrically equivalent to $-6$~T at $\psi=0^{\circ}$, the spectra for these two angles are identical. 
In all the way from $\psi=0^{\circ}$ to $180^{\circ}$, the intensity is always enhanced at 6.712~keV ($E2$) and suppressed at 6.720~keV ($E1$).

Next, the field was reversed to $-6$~T at $\psi=0^{\circ}$, and the energy scans were carried out with the rotation of the sample to $\psi=180^{\circ}$ by a step of 30$^{\circ}$. In this turn, the intensity is always suppressed at 6.712~keV ($E2$) and enhanced at 6.720~keV ($E1$).
These results show that the relationship between the atomic displacement (local expansion or contraction) and the magnitude of the Sm moment (large or small) along the field direction, which is reversed with the field reversal, does not change with the rotation of the sample in a constant field.

\subsection{Polarized neutron diffraction}
We next study by PND the relationship between the atomic displacement and the magnitude of the Sm moment. 
The diffraction intensity for a neutron beam with polarization $\bm{P}$, without analyzing the final polarization, is expressed as
\begin{align}
\Bigl( \frac{{\rm d}\sigma}{{\rm d}\Omega} \Bigr)
&= \bigl| F_{\text{N}} + r_0 \bm{F}_{\text{M}\perp}\cdot\bm{P} \bigr|^2 \,,
\label{eq:IntPND}
\end{align}
where $F_{\text{N}}$ and $\bm{F}_{\text{M}}$ are the nuclear and magnetic structure factors, respectively. 
$\bm{F}_{\text{M}\perp}$ represents the component perpendicular to the scattering vector. 
Since we already know $F_{\text{N}}$ from the previous x-ray diffraction study.\cite{Matsumura16}  
we can directly investigate $\bm{F}_{\text{M}}$ by using the nuclear--magnetic interference term. 

\begin{figure}[t]
\begin{center}
\includegraphics[width=8.5cm]{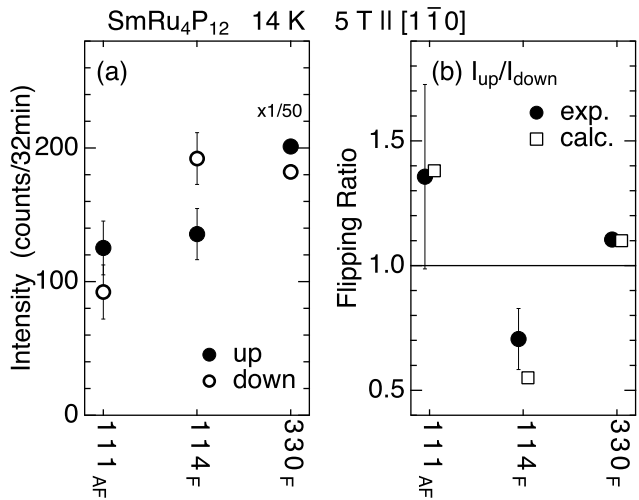}
\end{center}
\caption{(a) Intensities and (b) flipping ratios of the 111, 114, and 330 Bragg diffraction of polarized neutrons for up and down spin states in phase II at 5~T $\parallel [1 \bar{1} 0]$ and 14~K. 
The background has been subtracted. The flipping ratios are compared with the calculations. 
F and AF represent the ferro- and antiferro-magnetic Bragg points, respectively. 
}
\label{fig:PNDIntFR}
\end{figure}
Figure~\ref{fig:PNDIntFR}(a) shows the intensities of 111 (antiferromagnetic), 114 (ferromagnetic) and 330 (ferromagnetic) reflections for up and down spin polarizations. 
Note that a magnetic field of 5 T is applied upwards along the vertically oriented crystal axis of $[1 \bar{1} 0]$ with the $[110]-[001]$ horizontal scattering plane.  
The flipping ratios $I_{\text{up}}/I_{\text{down}}$ are shown in Fig.~\ref{fig:PNDIntFR}(b). 
To compare the flipping ratio with the calculation, we simply assume that the Sm moments at the corner (Sm-1) and the center (Sm-2) of the bcc unit cell may be expressed as 
$\bm{m}_1=\bm{m}_{\text{F}} +\bm{m}_{\text{AF}}$ and $\bm{m}_2=\bm{m}_{\text{F}} -\bm{m}_{\text{AF}}$, respectively, as shown in Fig.~\ref{fig:MagstModel}. 
From the magnetization data at 5~T and 14~K in phase II, we set $\bm{m}_{\text{F}} =(0.0424,-0.0424,0)$ $\mu_{\text{B}}$, which gives a uniform magnetization of $|\bm{m}_{\text{F}}|=0.06$~$\mu_{\text{B}}$.\cite{DKikuchi07} 
The calculated flipping ratios for 114 and 330 reflections are consistent with the observation. 

\begin{figure}[t]
\begin{center}
\includegraphics[width=7.5cm]{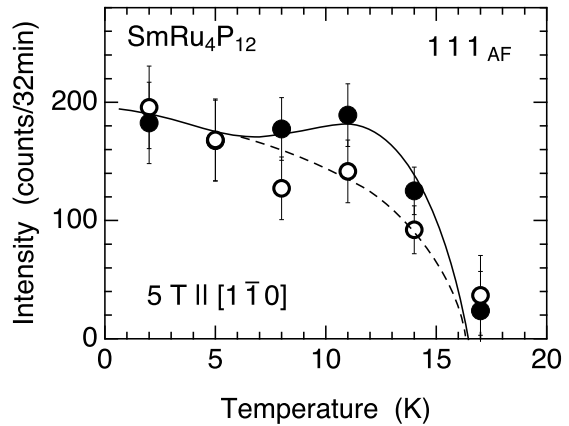}
\end{center}
\caption{Temperature dependence of the 111 antiferromagnetic Bragg intensities for up (solid circle) and down (open circle) spin polarizations. The background has been subtracted. The solid and broken lines are guides for the eye. }
\label{fig:PND111Tdep}
\end{figure}

Among the four $\langle 111 \rangle$ AFM domains, those with the moments parallel to the field are selected in phase II. 
We therefore set $\bm{m}_{\text{AF}}=(-0.0928,0.0928,0.0928)$~$\mu_{\text{B}}$. 
This gives an AF component of $|\bm{m}_{\text{AF}}|=0.16$~$\mu_{\text{B}}$. 
This is a reasonable assumption because the ordered moment is estimated to be 0.3~$\mu_{\text{B}}$ at the lowest temperature and the AF magnetic moment at 14~K is estimated from Fig.~\ref{fig:TdepFpsFpp110} to be about half the value at the lowest temperature.\cite{Lee12,Aoki07} 
This model gives a smaller moment of $|\bm{m}_1|=0.117$~$\mu_{\text{B}}$ for Sm-1 and a larger moment of $|\bm{m}_2|=0.213$~$\mu_{\text{B}}$ for Sm-2.  
By using the atomic displacement parameters obtained in Ref.~\onlinecite{Matsumura16} for $\bm{H}\parallel [110]$, corresponding to the case (a) in Fig.~\ref{fig:MagstModel}, we have the calculated flipping ratio for 111 as shown in Fig.~\ref{fig:PNDIntFR}(b), which well reproduces the experimental observation. 

Figure~\ref{fig:PND111Tdep} shows the $T$-dependence of the 111 Bragg intensity for up and down spin polarizations. 
We can see that the intensity for up spin becomes higher than that for down spin in phase II because of the appearance of $F_{\text{N}}$ in addition to $F_{\text{M}}$. 
As we analyzed above, these data provide a strong piece of evidence that the P$_{12}$ cage and Ru cube expand (contract) around the small (large) moment site of Sm. 
However, although these data of PND are direct and have no ambiguity, we must admit that the statistical error bar of the 111 intensity is still large to be conclusive. 
It should be confirmed by another method, which will be done later by analyzing the RXD data of Fig.~\ref{fig:Escan15KppH110}. 

\subsection{Resonant Soft X-ray Diffraction}
One of the important aims of this study is to directly detect the charge ordering of the $p$-electrons in the $a_{u}$ band consisting of the P$_{12}$ molecular orbitals. 
RSXD at the P $K$-edge, which directly observes the $3p$ state via the $1s$--$3p$ transition, is expected to be the best tool. 
We also expect that the $p\,$--$f$ hybridization would induce a spin polarization in the $3p$ orbital in the AFM phases II and III, which should be detected as a resonant scattering of magnetic origin.  
\begin{figure}[t]
\begin{center}
\includegraphics[width=8.5cm]{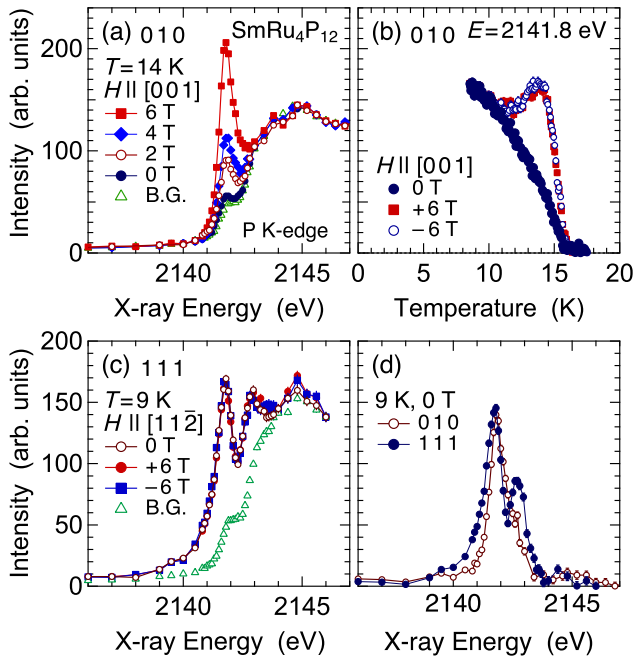}
\end{center}
\caption{Results of resonant soft x-ray diffraction at the P $K$-edge using a $\pi$-polarized incident beam without analyzing the final polarization.  
(a) X-ray energy dependences of the 010 Bragg intensity at 14~K in phase II in magnetic fields applied along the [001] axis. BG represents the background signal due to the fluorescence. 
(b) Temperature dependence of the 010 intensity at resonance in magnetic fields of 0 and $\pm 6$~T. The background has been subtracted. 
(c) X-ray energy dependences of the 111 Bragg intensity at 9~K in phase III in magnetic fields of 0 and $\pm 6$~T along the $[11\bar{2}]$ axis. 
(d) Comparison of the energy spectrum of the 010 and 111 Bragg intensities at 9~K at zero field after background subtraction.}
\label{fig:PKedge}
\end{figure}

Figure~\ref{fig:PKedge}(a) shows the energy spectrum of the 010 Bragg intensity in phase II in magnetic fields along the [001] axis. 
The resonant intensity at 2141.8~eV increases with increasing the field. Since we have not analyzed the polarization of the diffracted x-ray, we cannot conclude whether this increase reflect the development of the charge order or the increase in the magnetic $\pi$-$\pi'$ scattering. 
The $T$-dependence of this resonant intensity is shown in Fig.~\ref{fig:PKedge}(b). 
At zero field where only the AFM order exists, the resonant signal is purely of magnetic origin. The $T$-dependence is exactly the same as the one at the Sm $L_3$-edge. 
In magnetic fields, the intensity increases in phase II. It is noted that the intensity is equally enhanced for plus and minus field directions, which is different from the result at the Sm $L_3$-edge. 
Although there remains a possibility that this enhanced component is of charge origin (rank-0), or from the modified $p$ orbitals due to the lattice distortion (rank-2, Templeton-Templeton scattering), they both seems to be unlikely. If the resonant charge scattering were the case, it should behave in the same way as the non-resonant Thomson scattering and should interfere with the magnetic scattering, giving rise to the field-reversal asymmetry in intensity. The same should be the case also for the rank-2 scattering.
Therefore, we conclude that the enhanced intensity in phase II reflects the increase in magnetic scattering, i.e., the increase in the $\pi$-$\pi'$ scattering due to the enhancement of the parallel AFM component. 

The observation of the magnetic scattering in the P $K$-edge RSXD at a forbidden Bragg point directly shows that the P$_{12}$ molecular orbitals are spin polarized in a staggered manner. 
This is a direct evidence for the $p\,$--$f$ hybridization, which has been considered as being essential for the ordering phenomena in SmRu$_4$P$_{12}$.  

There are some points that should be remarked. 
First, the non-resonant Thomson scattering was not detected in RSXD. 
Although the reason is unclear, this must be the reason we did not observe the field-reversal asymmetry. 
Judging from the intensity observed in the Sm $L_3$-edge experiment, the non-resonant Thomson scattering should have been detected above the background level. 
Some surface effect, or the difference in the penetration depth, might be associated with. 
Second, the resonant intensity is much smaller than that of PrRu$_4$P$_{12}$, where the background intensity from the fluorescence is negligibly smaller than the signals of non-resonant Thomson and resonant scatterings.\cite{Nakao18,Nakao20} 
In PrRu$_4$P$_{12}$, the resonant signal was attributed mostly to the modified $p$-band by the staggered structural distortion.
The signal truly reflecting the $p\,$--$f$ hybridized state was considered to be much weaker. 
In this respect, here in SmRu$_4$P$_{12}$, where the non-resonant Thomson scattering was not detected, the resonant signal would be attributed to the spin polarized $p$-state through the $p\,$--$f$ hybridization. 
Finally, in the RSXD experiment performed at the Ru $L_3$-edge around 2840 eV ($2p_{3/2}\leftrightarrow 4d$), no signal was detected, indicating that the spin polarization of the $4d$-band through $d\,$--$f$ hybridization is weak, not playing a major role in the ordering phenomenon in SmRu$_4$P$_{12}$.

Figure~\ref{fig:PKedge}(c) shows the energy spectrum of the 111 Bragg intensity in phase III at the lowest temperature of 9~K. 
The intensity does not change by applying a magnetic field. This can be understood by considering that we do not analyze the final polarization, i.e., the $\pi$-$\pi'$ intensity decreases whereas the $\pi$-$\sigma'$ intensity increases due to the change in the domain population. 
Another important feature is that the resonant spectrum is apparently different from that of the 010 reflection.  
The resonant peak for 111 is wider than that for 010 and has a side peak at 2142.8~eV. 
The difference in the energy spectrum is shown in Fig.~\ref{fig:PKedge}(d). 
This anisotropy is also observed in PrRu$_4$P$_{12}$ and is attributed to the anisotropic $3p$ state of the P$_{12}$ molecular orbital.\cite{Nakao20}

\subsection{Determination of the magnetic spectral function}
\begin{figure}[t]
\begin{center}
\includegraphics[width=8cm]{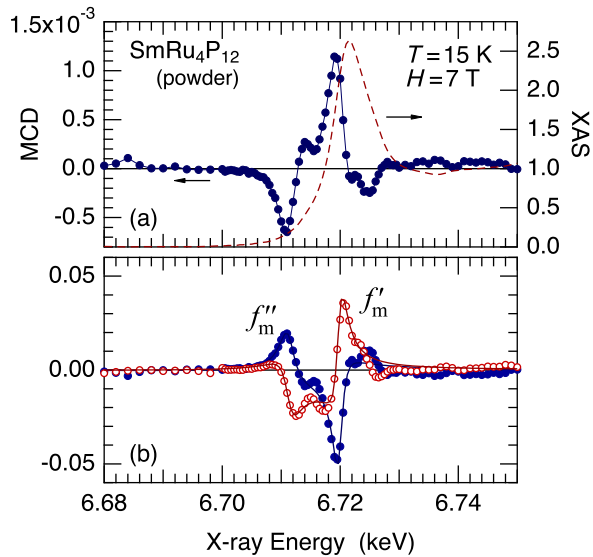}
\end{center}
\caption{(a) XMCD (filled circles) and XAS (dashed line) of SmRu$_4$P$_{12}$ at 15 K and 7 T in phase II around the Sm $L_3$-edge.  (b) Real and imaginary parts of the magnetic dipolar spectral function $f_{\text{m}}(\omega)$ in an arbitrary unit. 
Solid lines are the fits using Eq.~(\ref{eq:specfunc}).}
\label{fig:MCDfm}
\end{figure}

In order to deduce the relationship between $F_{\text{C}}$ and $F_{\text{M}}$ from the results of RXD, we need to determine the resonant spectral function $\alpha(\omega)$ for the magnetic dipole moment, i.e., the rank-1 spectral functions of $\alpha_{E2}^{(1)}(\omega)$ and $\alpha_{E1}^{(1)}(\omega)$. 
For this purpose, we have performed an XMCD measurement.  
As described in the Appendix, XAS and XMCD spectra are directly associated with the electric charge (rank-0) and magnetic dipolar (rank-1) spectral functions, respectively. 
Figure~\ref{fig:MCDfm}(a) shows the XMCD and XAS spectra at 15~K and 7~T in phase II.\cite{SM_L2} 
The XMCD spectrum, especially around the $E2$ resonance at 6.712~keV, is similar to those of other trivalent Sm compounds such as Sm$_2$Fe$_{14}$B, SmFe$_2$, and SmAl$_2$, indicating that the $E2$ transition ($2p\leftrightarrow 4f$) reflects mostly the atomic nature of Sm$^{3+}$.\cite{Parlebas06} 
The spectrum around the $E1$ resonance at 6.720~keV is similar to that of Sm$_2$Fe$_{14}$B, but is slightly different from those of SmFe$_2$ and SmAl$_2$. 
This would be because the $E1$ transition ($2p\leftrightarrow 5d$) is affected by the hybridization of the Sm-$5d$ with the surrounding orbitals.\cite{Harada04} 

From the XAS and XMCD spectra, we directly obtain the imaginary part $f''_{\text{m}}(\omega)$ by using (\ref{eq:A-13}). 
The real part $f'_{\text{m}}(\omega)$ is obtained by the Kramers-Kronig transformation. 
$f'_{\text{m}}(\omega)$ and $f''_{\text{m}}(\omega)$ thus obtained are shown in Fig.~\ref{fig:MCDfm}(b). 
To use these spectral functions in the analysis of the RXD data in Fig.~\ref{fig:Escan15KppH110}, we fit $f_{\text{m}}(\omega)$ by the following Lorentzian function:
\begin{align}
f_{\text{m}}(\omega) &= \alpha_{E2}^{(1)}(\omega) + \alpha_{E1}^{(1)}(\omega) \nonumber \\
&= \frac{I_2 \Gamma_2 e^{i\phi_2}}{\hbar\omega - \Delta_2 + i\Gamma_2}
+ \frac{ I_1 \Gamma_1 e^{i\phi_1}}{\hbar\omega - \Delta_1 + i\Gamma_1} \,.
\label{eq:specfunc}
\end{align}
The fitted curves for $f'_{\text{m}}(\omega)$ and $f''_{\text{m}}(\omega)$ are shown by the solid lines in Fig.~\ref{fig:MCDfm}(b). 
The phase parameters obtained are $\phi_2=-2.67$ rad and $\phi_1=0.61$ rad, which are close to the values we first tried in Sec.~III-B and not the ones shifted by $\pi$ rad. 

\begin{figure}[t]
\begin{center}
\includegraphics[width=8cm]{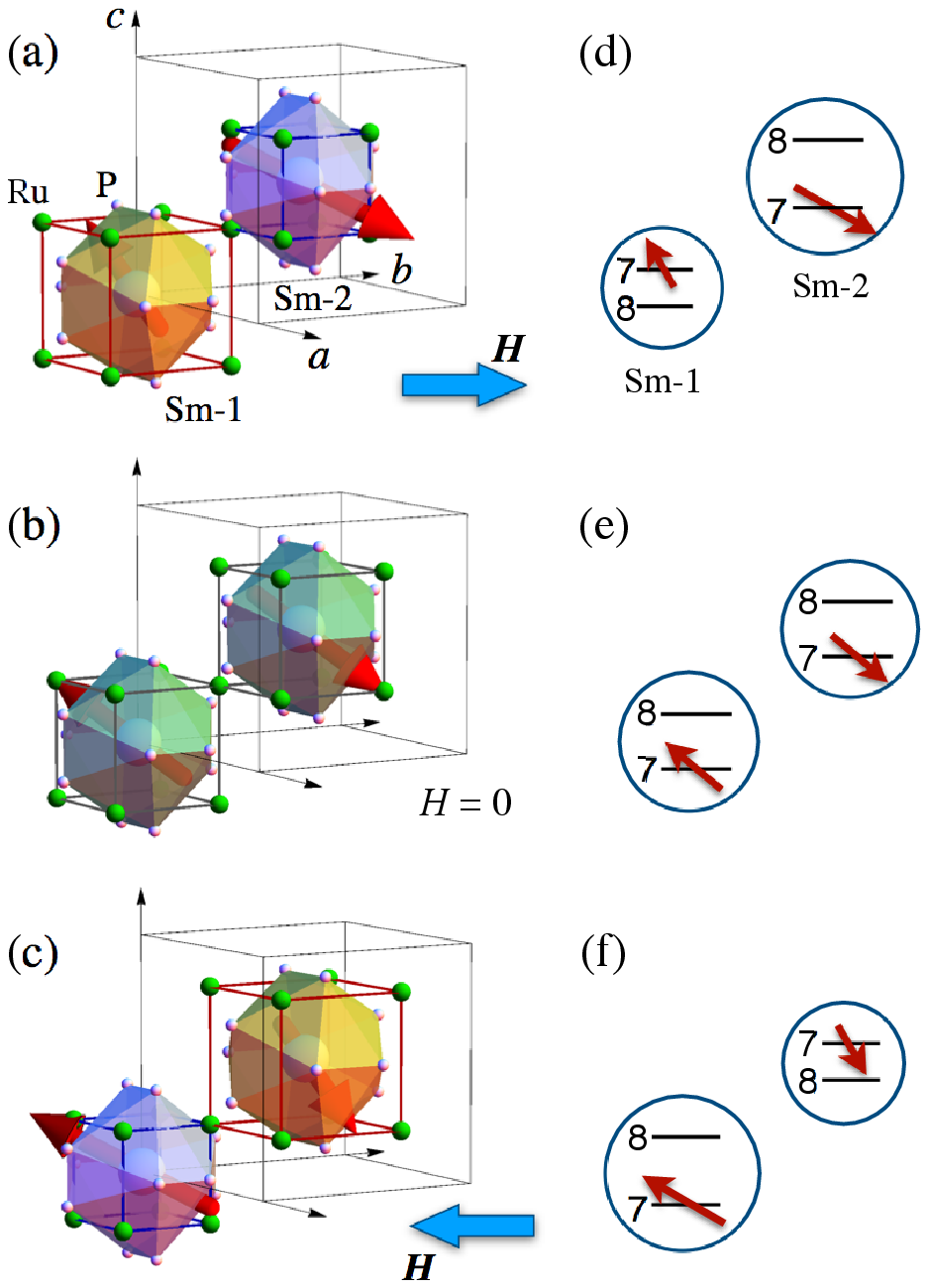}
\end{center}
\caption{(a) A model structure representing the relationship between the atomic displacements and the magnetic moments of Sm ions in magnetic fields applied along [110] in phase II. The surrounding cage of P$_{12}$ and Ru expand (contract) around the small (large) moment site of Sm-1 (Sm-2) where the AFM moment has an antiparallel (parallel) component along the applied field. 
(b) The AFM ordered state at zero field with equal moments at Sm-1 and Sm-2 and equal volumes of the surrounding cage. 
(c) The same as (a), but for the reversed field direction. The magnetic moment of Sm-1 becomes larger than that of Sm-2. 
(d), (e), (f) Schematic of the charge density of the $p$ electrons, which is represented by the size of the circle, and the electronic state of the $f$ electrons, corresponding to the situation of (a), (b), (c), respectively. 
The indices of 7 and 8 represent the $\Gamma_7$-like and $\Gamma_8$-like states, respectively. 
}
\label{fig:FigMagst}
\end{figure}

Using these parameters, we analyzed again the energy spectrum of RXD using Eq.~(\ref{eq:IntRXD}), the result of which is shown by the solid lines in Fig.~\ref{fig:Escan15KppH110}. 
The parameters obtained are $F_{\text{C}}=9.86(-0.331 + 0.0232i)$, using the exact phase for the case of local expansion around Sm-1, $I_{2}\bm{G}_{E2}^{(1)}\cdot\bm{F}_{\text{M}}=2.66$ and $I_{1}\bm{G}_{E1}^{(1)}\cdot\bm{F}_{\text{M}}=2.56$.  
We can now conclude that the case~(a) in Fig.~\ref{fig:MagstModel} is realized, 
i.e., the P$_{12}$ cage and the Ru cube expand (contract) around the small moment site of Sm where the AFM moment is antiparallel to the applied field. 
This is the same result as the one obtained from PND. 
Energy dependences of the 111, 223, 221, and 225 Bragg reflections, which have different structure factors of $F_{\text{C}}$, are also explained consistently.\cite{SM_111} 
Therefore, the results of PND and RXD now complementarily give conclusive evidence on the relationship between the atomic displacement and the magnitude of the Sm moment, which is shown in Fig.~\ref{fig:FigMagst}.

When the field direction is reversed from $[110]$ to $[\bar{1}\bar{1}0]$, the magnetic moment of Sm-1 becomes larger and that of Sm-2 becomes smaller as shown in Fig.~\ref{fig:FigMagst}(c) because they have parallel and antiparallel components along the field, respectively. 
It is noted that the magnetic structure factor $\bm{F}_{\text{M}}=\bm{m}_1 - \bm{m}_2$ does not change by the field reversal. 
However, since the local lattice distortion follow the change in the $4f$ and $p$ states, the cage of P$_{12}$ and Ru contract around Sm-1 and expand around Sm-2, which results in the change in the sign of $F_{\text{C}}$. 
This is the reason why the interference changes by reversing the field direction. 
On the other hand, when we rotate the crystal in a constant field, the interference does not change as described in Sec.~III-C. 
In Fig.~\ref{fig:FigMagst}(a), the large (small) moment of Sm-2 (Sm-1) keeps being oriented parallel (antiparallel) to the field throughout the rotation without changing its magnitude and the local volume of the surrounding cage. 
Thus, the relation between the magnetic moment and the lattice does not change.

\section{Discussion}
\subsection{The relation among the order parameters}
We have concluded that the P$_{12}$ cage and the Ru cube contract (expand) around the large (small) moment site of Sm. 
Intriguingly, this relation is opposite to the cases in PrRu$_4$P$_{12}$ and PrFe$_4$P$_{12}$.\cite{Iwasa05b,Iwasa12} 
Although it is difficult to give a clear explanation at the present stage, let us discuss this difference. 
In SmRu$_4$P$_{12}$ with $f^5$ ($J=5/2$) configuration, the $\Gamma_7$ doublet has a strong hybridization with the $a_u$ conduction band of the P$_{12}$ molecular orbitals.\cite{Shiina13}  
Consequently, when the staggered ordering of the CF states takes place without breaking the local symmetry, the Sm ion with the $\Gamma_7$-like ground state has larger magnetic moment than that with the $\Gamma_8$-like ground state. 
In contrast, in PrRu$_4$P$_{12}$, the nonmagnetic $\Gamma_1$ orbital has much stronger hybridization with the $a_u$ band than the magnetic $\Gamma_4^{(2)}$ orbital.\cite{Shiina10} 
That is, the CF state that has strong hybridization with the $a_u$ band has opposite magnetic character between Sm and Pr skutterudites. 

What is common in the charge ordered states in SmRu$_4$P$_{12}$ and PrRu$_4$P$_{12}$ is that the local volume of the P$_{12}$ cage and the Ru cube contracts at the rare-earth site with the ground state orbital possessing strong $p\,$--$f$ hybridization. 
According to a theoretical model for SmRu$_4$P$_{12}$, the $\Gamma_7$-like Sm sites with larger moments are expected to attract more $p$-electrons.\cite{Shiina16} 
This situation is schematically illustrated in Figs.~\ref{fig:FigMagst}(d) and (f). 
It is noted, however, the atomic displacements are secondary effects in response to the primary orderings of the charge density and the CF states. It is difficult to associate the atomic displacements with the primary order parameters in a straightforward manner without taking into account the total electronic energy involved in this change.  
If we simplistically consider only the $a_u$ molecular orbital, which has a relatively anti-bonding character, the increase in the charge density should lead to the expansion of the cage, which is opposite to our expectation. 
Here, it is useful to refer to a result of band calculation for PrRu$_4$P$_{12}$ taking into account the staggered displacement of the P atoms.\cite{Harima03} 
According to the calculation, the total number of electrons increase (including all the orbitals other than $a_u$) on the P atoms of the cage that contracts around Pr-2 at $(0.5, 0.5, 0.5)$.\cite{Harima20} 
This seems to be consistent with the picture of Figs.~\ref{fig:FigMagst}(d) and (f). 
In any case, although we have observed the spin polarization in the $p$ orbital by RSXD, we have no experimental evidence on the charge density. 
In future, it should be challenged to observe the difference in the charge densities of the two P$_{12}$ cages, which is estimated to be $\sim$1/100, i.e., the ratio of the energy scale of the charge order ($\sim$100 K) to the band width ($\sim$1 eV). 
It should also be challenged to observe the staggered ordering of the CF states of the two Sm sites that develops with increasing the magnetic field. 

When the field direction is reversed, the atomic displacements and the magnitude of the Sm moments are also reversed by conserving the relationship among them. 
This means that the order parameters are linearly and continuously modified by the applied field, which leads to the continuous modification of the ordered state in phase II when the sample is rotated and the field direction is changed. 
This must be basically the same phenomenon as the one observed in PrFe$_4$P$_{12}$ at low magnetic fields in the unusual nonmagnetic ordered phase where parallel AFM is always induced for any field dlrection.~\cite{JKikuchi07,Sakai07}

\subsection{Phase II to III transition}
The transition from phase II to III is relatively broad for $\bm{H} \parallel [100]$.\cite{Matsumura14}  
However, in the present experiment for $\bm{H} \parallel [110]$, the transition is sharp as we can see in Fig.~\ref{fig:TdepFpsFpp110}. 
This is also the case for $\bm{H} \parallel [112]$.\cite{SM2} 
This result is consistent with the sharp anomaly observed in the recent specific-heat measurement for $\bm{H} \parallel [111]$ and [110].\cite{Fushiya16} 
This anisotropy in the transition is considered to be related with the domain distribution. 
For $\bm{H} \parallel [100]$, the four domains with the AFM moments oriented along [111], $[1\bar{1}1]$, $[\bar{1}11]$, and $[\bar{1}\bar{1}1]$ are equivalent both in phase II and III. 
Domain population does not change by the transition between phase II and III. 
The change in the magnetic moment is expected to be small and the phase transition looks like a crossover. 

In contrast, for $\bm{H} \parallel [110]$ or $[112]$, as reported in Ref.~\onlinecite{Matsumura16}, different domains are selected in phases II and III. 
In phase III, the AFM moments prefer to be perpendicular to the field, whereas in phase II they prefer to be parallel to the field. Therefore, for $\bm{H} \parallel [110]$, the [111] and $[\bar{1}\bar{1}1]$ domains are selected in phase II, and change to $[\bar{1}11]$ and $[1\bar{1}1]$ domains in phase III. 
Since the change in the direction of the ordered moments accompany changes in the $f$-electron state, the phase transition for $\bm{H} \parallel [110]$ is expected to experience a large change in entropy, leading to a sharp anomaly in the order parameters.

\section{Summary}
We have performed resonant x-ray diffraction (RXD) to clarify the nature of the field-induced charge ordered phase (phase II) of SmRu$_4$P$_{12}$, especially to clarify the relationship between the magnitudes of the Sm moments and the atomic displacements of P and Ru surrounding the Sm atom. 
The P$_{12}$ cage and the Ru cube contract (expand) around the large (small) moment site of Sm.
X-ray magnetic circular dichroism was utilized to determine the rank-1 spectral function for the magnetic dipole moment, with which we concluded the above relation from RXD by analyzing the interference between the Thomson scattering from the atomic displacements and the resonant scattering from the AFM order. 
We also performed a flipping-ratio measurement in polarized neutron diffraction to obtain complementary and unambiguous conclusion. 
Based on this result, we discussed the details of the orbital dependent $p\,$--$f$ hybridization responsible for this intriguing phase transition. It is remarked that this relation is opposite to the cases in PrRu$_4$P$_{12}$ and PrFe$_4$P$_{12}$. 

Isotropic and continuous nature of phase II was demonstrated by showing that the interference spectrum of RXD is invariant to the field direction. 
In phase II, a magnetic domain is always selected so that the AFM component becomes parallel to the applied field and gives rise to large and small moments, which is associated with the formation of the staggered ordering of the $\Gamma_7$-like and $\Gamma_8$-like CF states. 

In resonant soft x-ray diffraction (RSXD) at the P $K$-edge ($1s\leftrightarrow 3p$), we detected a resonance due to the spin polarized $3p$ orbitals reflecting the AFM order of Sm. We could not observe resonance, on the other hand, at energies around the Ru $L_3$-edge ($2p\leftrightarrow 4d$). 
This result shows that the $p\,$--$f$ hybridization indeed plays more important role than the $d\,$--$f$ hybridization in the exchange interaction in SmRu$_4$P$_{12}$. 
The charge order of the $p$ electrons in the P$_{12}$ molecular orbitals was not detected by RSXD probably because the modulation in the charge density was too small to be detected.

\acknowledgments
The authors acknowledge valuable discussions with R. Shiina and H. Harima. 
This work was supported by JSPS KAKENHI Grant Numbers JP20102005 (Heavy Electrons), JP15K05175, JP15H05885 (J-Physics), JP17K05130, JP17H05209, JP18H01182, JP18K18737, and JP20H01854. 
The synchrotron radiation experiments at SPring-8 were performed under Proposal Nos. 2015A3711 (BL22XU) and 2019B1247 (BL39XU). 
The synchrotron radiation experiments at KEK was performed under the approval of the Photon Factory Program Advisory Committee (Proposal Nos. 2012S2-005, 2017PF-BL-19B, and 2017G553). 
The neutron scattering experiment at ORNL was supported by the U.S.--Japan Cooperative Program on Neutron Scattering. 
This research used resource at the High Flux Isotope Reactor, a DOE Office of Science User Facility operated by the Oak Ridge National Laboratory.

\appendix*
\section{Formalism of resonant x-ray diffraction and x-ray magnetic circular dichroism}
\subsection{Resonant x-ray diffraction}
The intensity of x-ray diffraction is proportional to the square of the total structure factor, which is expressed as
\begin{align}
F(\bm{\kappa}, \omega) &= \sum_{j} f_{j} (\bm{\kappa}, \omega) e^{-i \bm{\kappa} \cdot \bm{r}_j}\,,
\label{eq:A-1}
\end{align}
where $\bm{\kappa} = \bm{k}'-\bm{k}$ represents the scattering vector and $\hbar\omega$ the photon energy. 
$f_{j} (\bm{\kappa}, \omega)$ is the energy-dependent atomic scattering factor of the $j$th atom at $\bm{r}_j$,  
which is generally expressed as
\begin{align}
f(\bm{\kappa}, \omega) &= f_{\text{nr}}(\bm{\kappa}) + \sum_{\nu=0}^{2} f_{E1}^{(\nu)}(\omega) + \sum_{\nu=0}^{4} f_{E2}^{(\nu)}(\omega) \;.
\label{eq:A-2}
\end{align}
The subscript $j$ is omitted hereafter. The first term represents the non-resonant Thomson scattering from all the electric charges of the atom:
\begin{align}
 f_{\text{nr}}(\bm{\kappa}) &= f_{0}(\bm{\kappa}) [\bm{\varepsilon}^{\prime *} \cdot \bm{\varepsilon} ] \,.
\label{eq:A-3}
\end{align}
$\bm{\varepsilon}$ and $\bm{\varepsilon}^{\prime}$ are the polarization vectors of the incident and scattered x-rays, respectively. 
The second and third terms represent the resonant scattering factors due to electric dipole ($E1$) and electric quadrupole ($E2$) transitions, respectively, which should be taken into account in the vicinity of an absorption edge of some specific element. 
The $E1$ and $E2$ scattering factors have sensitivities up to rank-2 (electric quadrupole) and rank-4 (electric hexadecapole) moments, respectively, as expressed by the summation over $\nu$.\cite{Lovesey05} 

The resonant scattering factor consists of the spectral function $\alpha^{(\nu)}(\omega)$ and the geometrical factor $\bm{G}^{(\nu)}$, which are both rank dependent.\cite{Nagao10} 
The $E1$ scattering factors for rank-0 (electric charge) and rank-1 (magnetic dipole) moments are expressed as 
\begin{align}
f_{E1}^{(0)}(\omega) &= \alpha_{E1}^{(0)}(\omega) [\bm{\varepsilon}^{\prime *} \cdot \bm{\varepsilon} ] \,,
\label{eq:A-4}  \\
f_{E1}^{(1)}(\omega) &= i \, \alpha_{E1}^{(0)}(\omega) [\bm{\varepsilon}^{\prime *} \times \bm{\varepsilon} ] \cdot \bm{m} \,,
\label{eq:A-5}
\end{align}
and for $E2$, 
\begin{align}
f_{E2}^{(0)}(\omega) &= \alpha_{E2}^{(0)}(\omega) [ 
(\bm{\varepsilon}^{\prime *} \cdot \bm{\varepsilon}) (\hat{\bm{k}}' \cdot \hat{\bm{k}})  \nonumber \\
& \;\; +(\bm{\varepsilon}^{\prime *} \cdot \hat{\bm{k}} ) (\hat{\bm{k}}' \cdot \bm{\varepsilon}) ] \,,
\label{eq:A-6} \\
f_{E2}^{(1)}(\omega) &= i \, \alpha_{E2}^{(1)}(\omega) [ 
(\bm{\varepsilon}^{\prime *} \cdot \bm{\varepsilon}) (\hat{\bm{k}}' \times \hat{\bm{k}}) \nonumber \\
& \;\; +(\bm{\varepsilon}^{\prime *} \times \bm{\varepsilon}) (\hat{\bm{k}}' \cdot \hat{\bm{k}}) 
 + (\hat{\bm{k}}' \cdot \bm{\varepsilon}) (\bm{\varepsilon}^{\prime *} \times \hat{\bm{k}} ) \nonumber \\
& \;\; + (\bm{\varepsilon}^{\prime *} \cdot \hat{\bm{k}} )(\hat{\bm{k}}' \times \bm{\varepsilon}) ] \cdot \bm{m} \,,
\label{eq:A-7}
\end{align}
where $\bm{m}$ represents the magnetic dipole moment of the atom under consideration. 
The terms in the square brackets from (\ref{eq:A-3}) to (\ref{eq:A-7}) are the geometrical factors $\bm{G}^{(\nu)}$, which have different forms depending on the respective scattering processes.

\subsection{X-ray magnetic circular dichroism}
For the forward scattering with $\bm{k}'=\bm{k}$, the $E2$ scattering factors of (\ref{eq:A-6}) and (\ref{eq:A-7}) become 
\begin{align}
f_{E2}^{(0)}(\omega) &= \alpha_{E2}^{(0)}(\omega)  [\bm{\varepsilon}^{\prime *} \cdot \bm{\varepsilon}] \,,
\label{eq:A-8} \\
f_{E2}^{(1)}(\omega) &= i \, \alpha_{E2}^{(1)}(\omega)  [\bm{\varepsilon}^{\prime *} \times \bm{\varepsilon}]  \cdot \bm{m}  \,,
\label{eq:A-9}
\end{align}
the same form as the $E1$ scattering factors of (\ref{eq:A-4}) and (\ref{eq:A-5}). 
Then, by rewriting $\alpha_{E2}^{(0)}(\omega)+\alpha_{E1}^{(0)}(\omega)= f_0'(\omega) + i f_0''(\omega)$ and 
$\alpha_{E2}^{(1)}(\omega)+\alpha_{E1}^{(1)}(\omega)= f'_{\text{m}}(\omega) + if''_{\text{m}}(\omega)$, 
we can rewrite (\ref{eq:A-2}) for the forward scattering as
\begin{align}
f(\omega) &= \{ f_{0} + f_0'(\omega) + i f_0''(\omega) \} [ \bm{\varepsilon}^{\prime *} \cdot \bm{\varepsilon} ] \nonumber \\
&\;\; + i\, \{ f'_{\text{m}}(\omega) + if''_{\text{m}}(\omega) \} [ \bm{\varepsilon}^{\prime *} \times \bm{\varepsilon} ] \cdot \bm{m} \,.
\label{eq:A-10}
\end{align}

The polarization vector of the circularly polarized x-ray with $\pm$ helicity is written as 
$\bm{\varepsilon}_{\pm} = (\bm{\varepsilon}_{\sigma} \pm i \bm{\varepsilon}_{\pi})/\sqrt{2}$, where we define $\bm{\varepsilon}_{\sigma}$ and $\bm{\varepsilon}_{\pi}$ so that the relation $\bm{\varepsilon}_{\sigma} \times \bm{\varepsilon}_{\pi} \parallel \hat{\bm{k}}$ is satisfied. 
We also use a description in which the electric field of an electromagnetic wave is expressed as $\bm{E}(\bm{r},t) \propto \bm{\varepsilon} e^{i(\bm{k}\cdot\bm{r} - \omega t)}$.

When we write the absorption coefficient for a circularly polarized x-ray with $\pm$ helicity as $\mu_{\pm}=\mu_0 \pm \Delta \mu$, the average and difference of $\mu_{\pm}$ become proportional to the XAS and XMCD spectrum, respectively: 
\begin{align}
\mu_{+} + \mu_{-}   &= - a  f_0''(\omega)  = 2\mu_0 \,, \label{eq:A-11} \\
\mu_{+} - \mu_{-}   &=  a (\hat{\bm{k}}\cdot \bm{m})  f''_{\text{m}}(\omega) = 2 \Delta \mu \,, \label{eq:A-12}  \\
\frac{\mu_{+} - \mu_{-}}{\mu_{+} + \mu_{-}} &= - \frac{ (\hat{\bm{k}} \cdot \bm{m}) f''_{\text{m}}(\omega)  }{f_0''(\omega)} \,.\label{eq:A-13}
\end{align}
$a=4 r_{\text{e}} \lambda / v_{\text{c}}$ is a constant factor, where $r_{\text{e}}$ represents a classical electron radius, $\lambda$ the wavelength of the x-ray, and $v_{\text{c}}$ the unit cell volume. 
We normally plot the XMCD spectrum for the geometry with $\hat{\bm{k}} \cdot \bm{m} < 0$ ($\bm{H} \parallel -\bm{k}$).\cite{Parlebas06} 
From the experimental results of XAS and XMCD, we can deduce $f''_{\text{m}}(\omega)$ using Eq.~(\ref{eq:A-13}). 
By applying a Kramers-Kronig transformation to $f''_{\text{m}}(\omega)$, we obtain $f'_{\text{m}}(\omega)$.


\clearpage
\begin{center}
\textbf{Supplemental Material}
\end{center}
\setcounter{section}{0}
\setcounter{equation}{0}
\setcounter{figure}{0}
\setcounter{table}{0}
\setcounter{page}{1}	
\makeatletter
\renewcommand{\theequation}{S\arabic{equation}}
\renewcommand{\thefigure}{S\arabic{figure}}
\renewcommand{\thetable}{S-\Roman{table}}
\renewcommand{\bibnumfmt}[1]{[S#1]}
\renewcommand{\citenumfont}[1]{S#1}

\section*{$\bm{H \parallel [\bar{1}\bar{1}2]}$}
\subsection{Domain population in phase III}
In the main text we described on the results for $H \parallel [1\bar{1}0]$. 
We show here the results for $H \parallel [\bar{1}\bar{1}2]$, which are almost the same as those for $H \parallel [1\bar{1}0]$. 

\begin{figure}[b]
\begin{center}
\includegraphics[width=8.5cm]{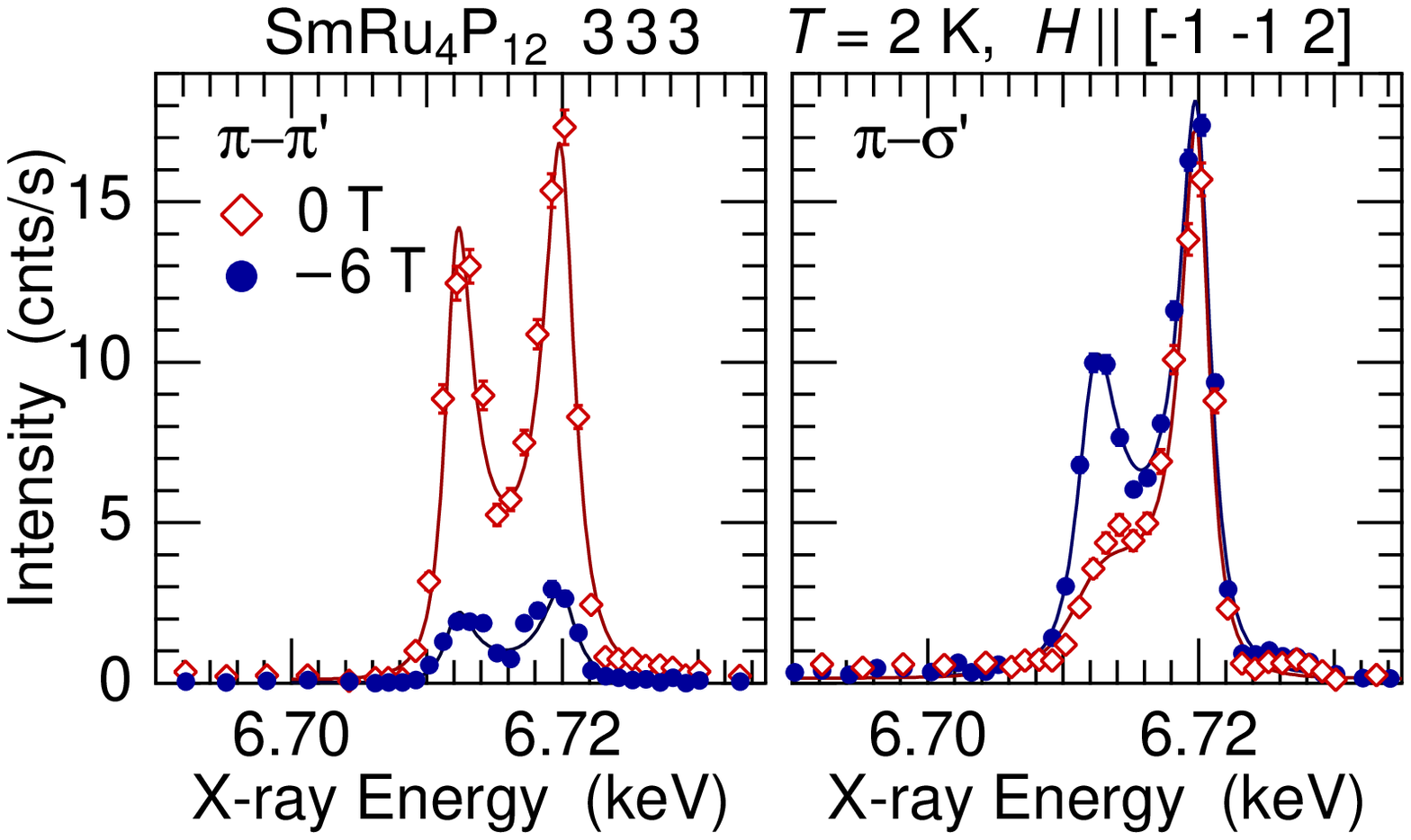}
\end{center}
\caption{X-ray energy dependences of the intensity of the 333 Bragg diffraction at 2 K in magnetic fields of 0 and $\pm 6$ T along  $[\bar{1}\bar{1}2]$. After absorption correction.}
\label{fig:Edep2KH112}
\end{figure}

Figure~\ref{fig:Edep2KH112} shows the resonance spectra of the 333 Bragg diffraction at 2 K in magnetic fields of 0 and $\pm 6$ T along  $[\bar{1}\bar{1}2]$. 
As described in the main text for $H \parallel [1\bar{1}0]$, the $\pi$-$\pi'$ intensity decreases significantly by applying the field. 
This shows that the volume fraction of the AFM domain increases in which the moments are oriented along $[111]$, perpendicular to the applied field ($\parallel$ the horizontal scattering plane), and give rise to the $\pi$-$\sigma'$ scattering according to Eqs.~(A.5) and (A.7). 
In contrast, the volume fractions of other AFM domains, which have parallel components to the field, decrease by applying the field, resulting in the decrease in the $\pi$-$\pi'$ intensity. 

\begin{figure}[t]
\begin{center}
\includegraphics[width=8.5cm]{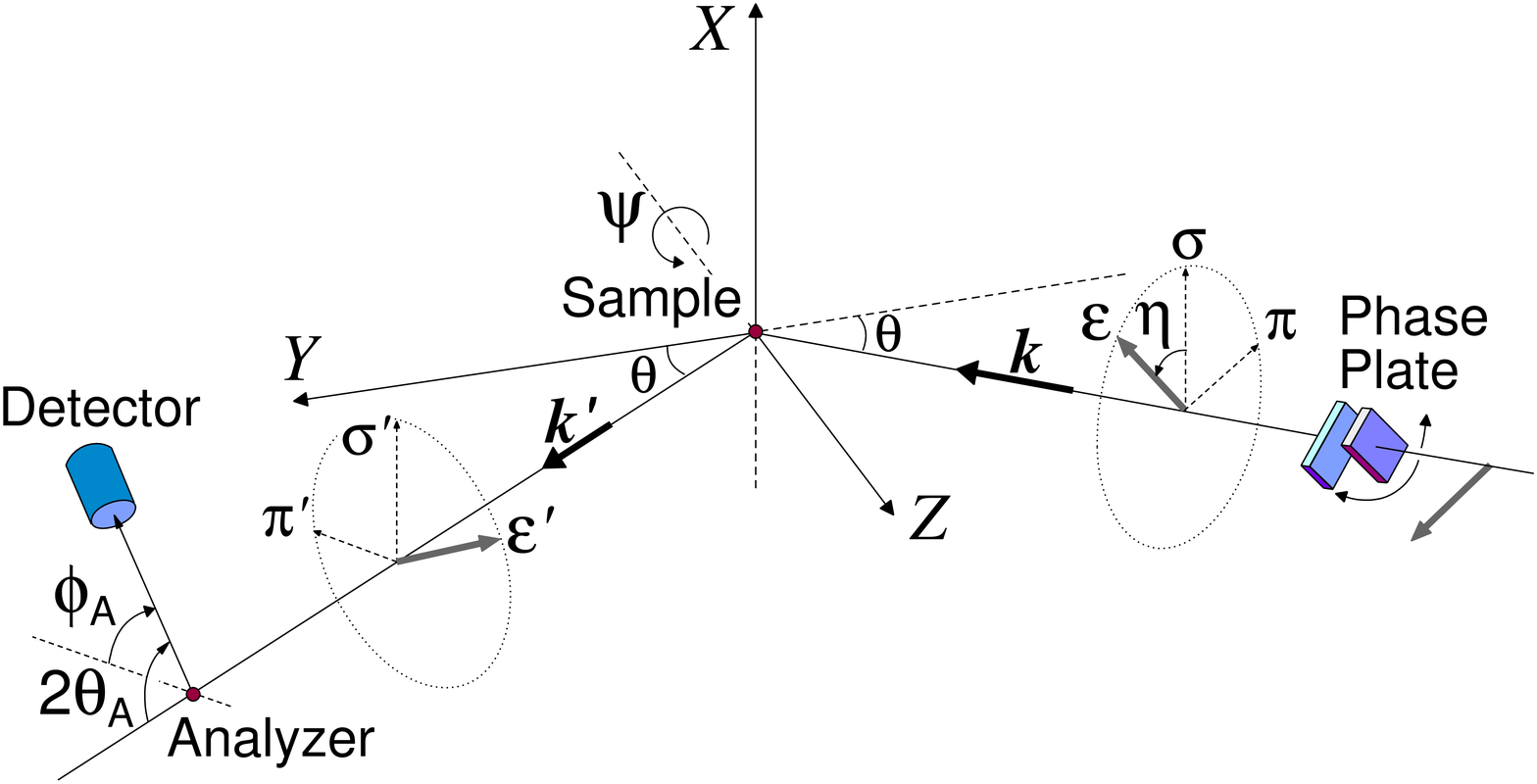}
\end{center}
\caption{Scattering geometry of the full-linear-polarization-analysis experiment. The polarization of the scattered x-ray is analyzed by rotating the analyzer crystal by an angle $\phi_{\text{A}}$. 
At $\phi_{\text{A}}=0^{\circ}$, only the $\sigma'$-polarization is detected by the analyzer, whereas at $\phi_{\text{A}}=90^{\circ}$ only the $\pi'$-polarization is detected. Linear polarization of the incident x-ray is manipulated by the double phase retarder system [34]. }
\label{fig:scattconfig}
\end{figure}

\begin{figure}[b]
\begin{center}
\includegraphics[width=8.5cm]{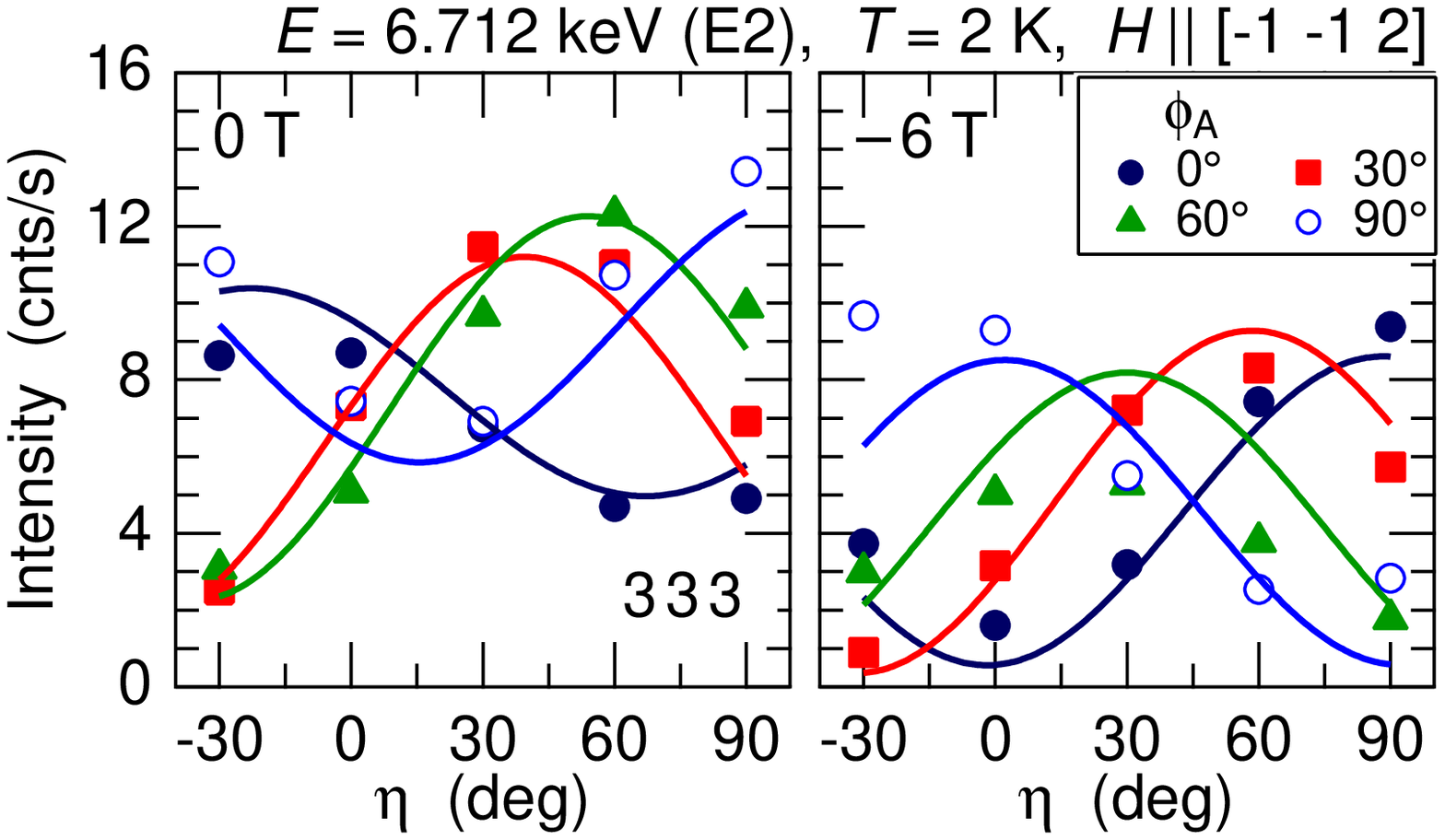}
\end{center}
\caption{Full-linear-polarization-analysis of the 333 Bragg diffraction at the $E2$ resonance at 2 K (phase III) in magnetic fields of 0 and $- 6$ T along  $[\bar{1}\bar{1}2]$. The definitions of the analyzer angle $\phi_{\text{A}}$ and the incident polarization angle $\eta$ are shown in Fig.~\ref{fig:scattconfig}. The solid lines are the calculations. }
\label{fig:Domain112E2}
\end{figure}

To investigate the domain population in more detail in phase III, we have performed a full linear polarization analysis by measuring the intensity as a function of the incident linear-polarization angle $\eta$ for several analyzer angles $\phi_{\text{A}}$, which are defined as shown in Fig.~\ref{fig:scattconfig}. 
In Fig.~\ref{fig:Domain112E2}, we show the data for the 333 reflection at the $E2$ resonance at 2 K in magnetic fields of 0 and $- 6$ T. The data are compared with the calculated curves assuming the magnetic dipolar geometrical factor for the $E2$ resonance. 
At 0 T, the data are well explained by assuming a population ratio of $0.20 : 0.20 : 0.13 : 0.47$ for the $[111]$, $[\bar{1}11]$, $[1\bar{1}1]$, and $[\bar{1}\bar{1}1]$ domains, respectively. 
The deviation from the equal population ratio of 0.25 could be due to some surface strain. 
At $-6$ T, in contrast, the calculated curves assume only the $[111]$ domain ($\bm{m}_{\text{AF}} \perp H$), which well explain the data and indicate the [111] domain is dominant.  
This shows that the phase III is a normal AFM phase in which the ordered moments prefer to be perpendicular to the field. 


\newpage
\subsection{Absence of $\bm{\sigma}$-$\bm{\sigma'}$ scattering in the $\bm{E1}$ resonance}
\begin{figure}[b]
\begin{center}
\includegraphics[width=8cm]{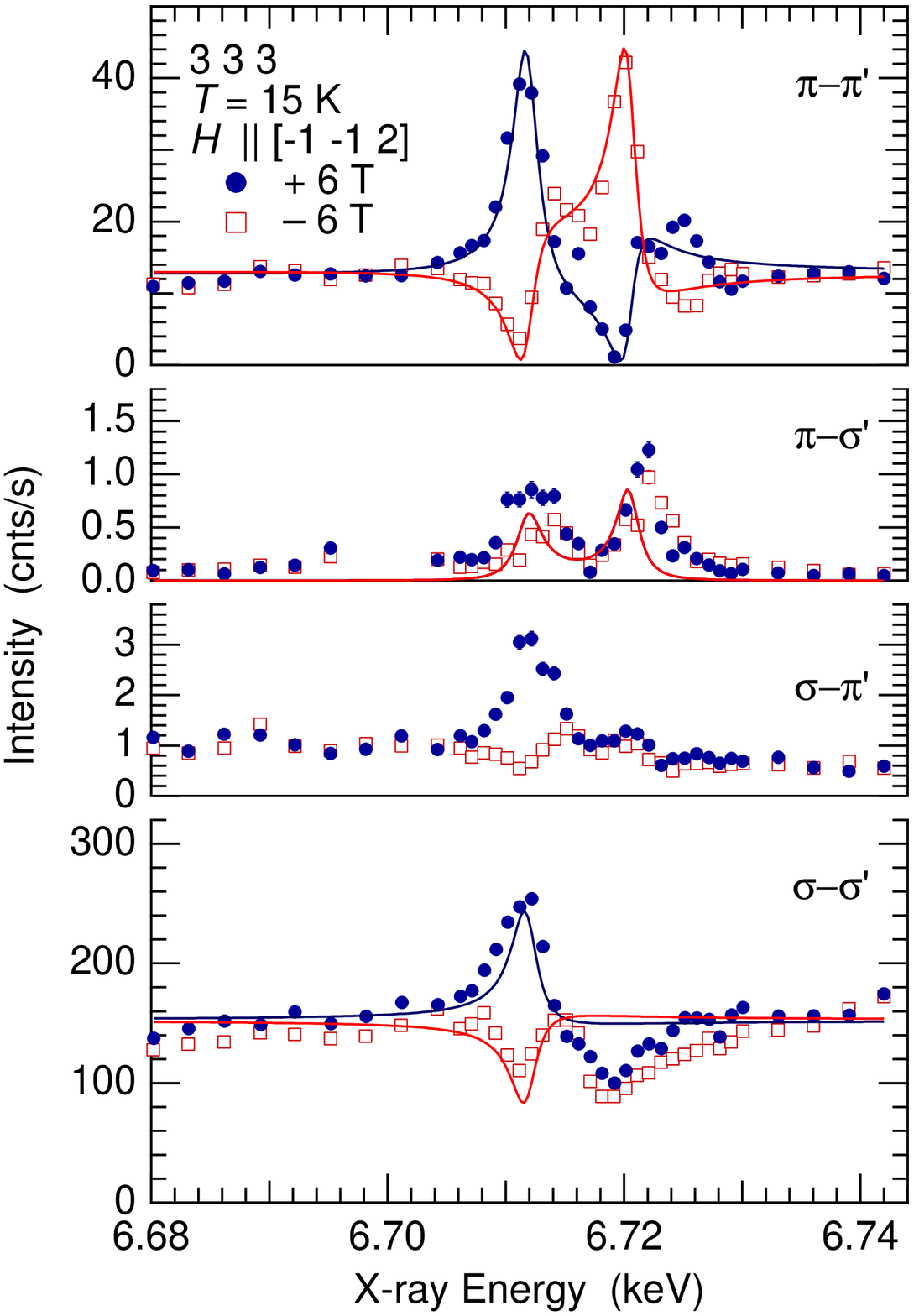}
\end{center}
\caption{X-ray energy dependences of the 333 Bragg-diffraction intensity for $\pi$-$\pi'$, $\pi$-$\sigma'$, $\sigma$-$\pi'$, and $\sigma$-$\sigma'$ scattering processes in magnetic fields of $\pm 6$ T at 15 K. 
After absorption correction. 
The solid lines are the fits to the data with Eq. (1) in the main text. 
}
\label{fig:Escan15KallpolsH112}
\end{figure}

Figure~\ref{fig:Escan15KallpolsH112} shows the resonance spectra of the 333 Bragg diffraction for the four scattering channels of $\pi$-$\pi'$, $\pi$-$\sigma'$, $\sigma$-$\pi'$, and $\sigma$-$\sigma'$. 
The purpose of this measurement is to show that the $\sigma$-$\sigma'$ scattering vanishes in the $E1$ resonance, which is expected for the resonant scattering of magnetic dipole origin. 
This can be observed in the bottom panel of Fig.~\ref{fig:Escan15KallpolsH112}. 
The field-reversal asymmetry at 6.720 keV ($E1$) is absent for $\sigma$-$\sigma'$. 
The dip in intensity around 6.720 keV is due to the imperfect correction of absorption at the main edge, which becomes  difficult for strong Bragg diffraction. 

The solid lines are the fits to the data with Eq. (1) in the main text, using the same phase factors of $\phi_2$ and $\phi_1$ determined from XMCD. 
The amplitudes of $F_{\text{C}}$, $I_2 F_{\text{M}}$, and $I_1 F_{\text{M}}$ were treated as free parameters. $F_{\text{C}}$ and $I_1$ were set to be zero for $\pi$-$\sigma'$ and $\sigma$-$\sigma'$, respectively. 
The intensities for $\sigma$-$\pi'$ are mostly due to the contamination from the $\sigma$-$\sigma'$ channel caused by the imperfect analyzer condition of $2\theta_{\text{A}}\neq 90^{\circ}$.  

\subsection{On the Rank-3 contribution to the $E2$ resonance}
There arises an $E2$ resonance and causes the interference with the Thomson scattering. 
This $E2$ resonance is of magnetic dipole origin which can be understood by the geometrical factor in Eq. (A.7) described in the main text. 
With respect to the possibility of a magnetic octupolar (rank-3) contribution to the $E2$ resonance, it is estimated to be weak and may be neglected because the polarization dependences shown in Fig.~\ref{fig:Domain112E2} are well explained by assuming the rank-1 (dipolar) geometrical factor as demonstrated by the calculated curves. 

\clearpage
\subsection{Temperature dependence}
\begin{figure}[b]
\begin{center}
\includegraphics[width=7.5cm]{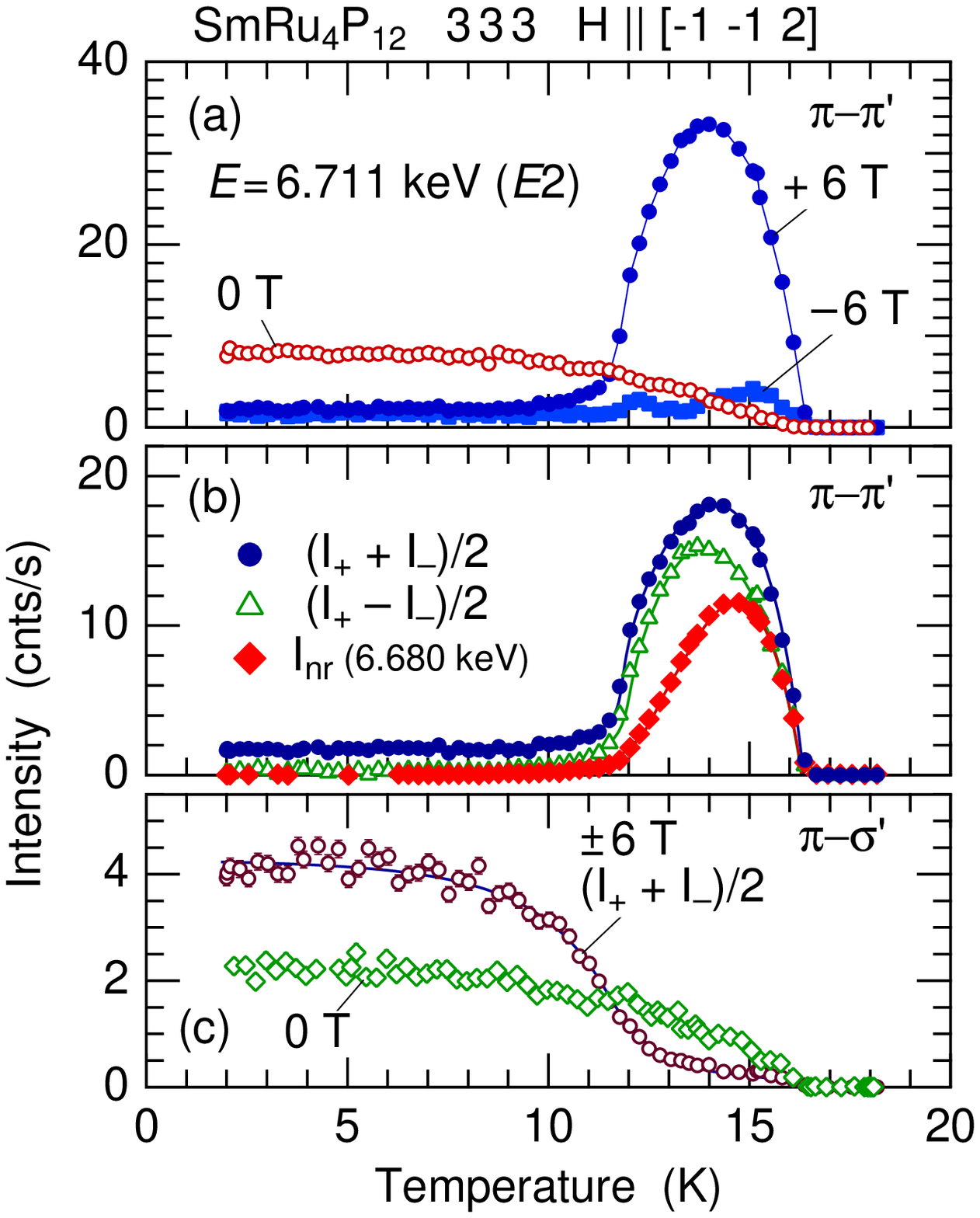}
\end{center}
\caption{(a) Temperature dependences of the 333 Bragg diffraction intensity for $\pi$-$\pi'$ at $E=6.712$ keV ($E2$) in magnetic fields of 0 and $\pm 6$ T. 
(b) Temperature dependences of the average and difference intensities for the data at $\pm 6$ T and the nonresonant intensity at 6.680 keV measured in the $\pi$-$\pi'$ channel. 
(c) Temperature dependences of the intensity at 0 T and the averaged intensity at $\pm 6$ T for $\pi$-$\sigma'$.  
 The lines are guides for the eye. }
\label{fig:TdepIntppH112}
\end{figure}

Figure~\ref{fig:TdepIntppH112} shows the temperature dependences of the $\pi$-$\pi'$ and $\pi$-$\sigma'$ intensities at the $E2$ resonance at zero field and $\pm 6$ T. 
The results are almost the same as those for $H \parallel [1\bar{1}0]$ in the main text. 

At zero field, both $\pi$-$\pi'$ and $\pi$-$\sigma'$ intensities exhibit a normal $T$-dependence of an AFM order. 
In magnetic fields, the $\pi$-$\pi'$ scattering exhibits a strong asymmetry with the field reversal in phase II. 
The $T$-dependence of the intensity of the non-resonant term is shown as $I_{\text{nr}}$, which has been measured at 6.680 keV and 6 T. This is proportional to $|F_{\text{C},\pi\pi'}|^2$. 
The development of the parallel AFM component on entering phase II is clearly detected as the enhancement of the average intensity $(I_{+} + I_{-})/2$, reflecting $|F_{\text{M},\pi\pi'}|^2$. 
The difference in intensity $(I_{+} - I_{-})/2$ shows the $T$-dependent interference effect, which is proportional to $F_{\text{C},\pi\pi'}F_{\text{M},\pi\pi'}$. 
In the $\pi$-$\sigma'$ channel, on the other hand, there arise little asymmetry in the field reversal and only the averaged intensity is shown. 
It is noted, however, the $\pi$-$\sigma'$ intensity, reflecting the perpendicular AFM component, is suppressed in phase II. 
This result consistently shows that the parallel AFM component is enhanced in phase II.

\begin{figure}[b]
\begin{center}
\includegraphics[width=7.5cm]{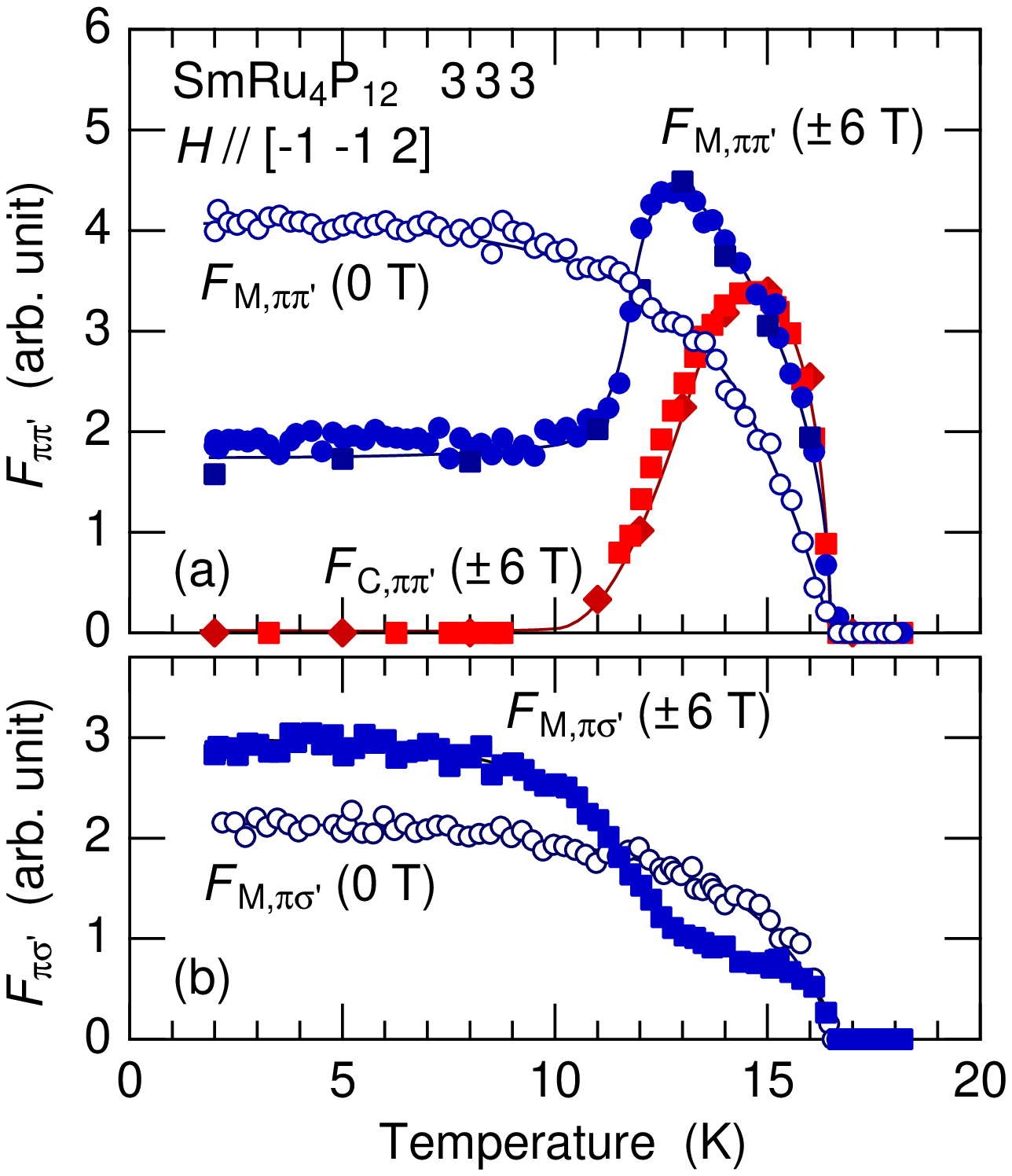}
\end{center}
\caption{Temperature dependences of the crystal and magnetic structure factors for $\pi$-$\pi'$ (top) and $\pi$-$\sigma'$ (bottom), which are deduced from the intensity in Fig.~\ref{fig:TdepIntppH112}. The solid lines are guides for the eye. }
\label{fig:TdepFpsFpp112}
\end{figure}
The $T$-dependences of the crystal and magnetic structure factors deduced from the intensity data are shown in Fig.~\ref{fig:TdepFpsFpp112}, which directly illustrates the behavior of the order parameters. 
The increase in $F_{\text{M},\pi\pi'}$ and the decrease in $F_{\text{M},\pi\sigma'}$ in phase II show that the $[\bar{1}\bar{1}1]$ AFM domain ($\parallel H$) is selected. 
The $[111]$ AFM domain ($\perp H$) is selected in the low temperature AFM phase III. 

\clearpage
\section*{111, 221, 223, and 225 reflections for $\bm{H \parallel [1\bar{1}0]}$}
\begin{figure}[b]
\begin{center}
\includegraphics[width=8.5cm]{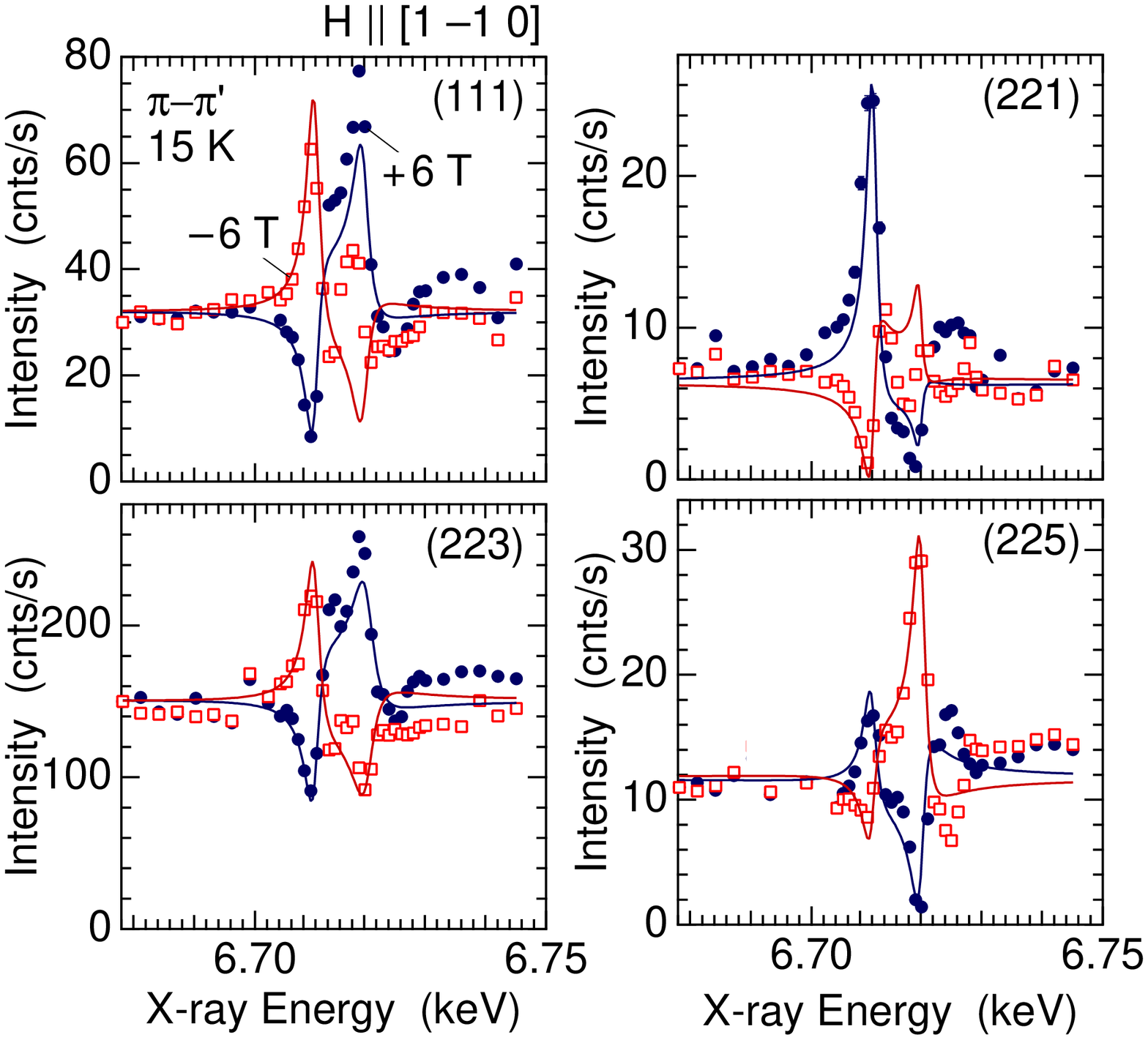}
\end{center}
\caption{X-ray energy dependences of the 111, 221, 223, and 225 Bragg intensities for the $\pi$-$\pi'$ channel at 15 K in magnetic fields of $\pm 6$ T along $[1\bar{1}0]$. After absorption correction. Solid lines are the fits with Eq.~(1) using the magnetic spectral function $\alpha^{(1)}(\omega)$ obtained from XMCD. }
\label{fig:Edep15KH110All}
\end{figure}

Figure~\ref{fig:Edep15KH110All} shows the x-ray energy dependences of some Bragg-diffraction intensities other than  333 reflection described in the main text. The fitting curves for each reflection were obtained in the same way as those for the 333 reflection explained in the main text. The same magnetic spectral function $f_{\text{m}}(\omega)$ obtained from XMCD is used. 
The interference anomalies of the 111 and 223 reflections are opposite to those of the 333, 221, and 225 reflections because the phase of the crystal structure factor $F_{\text{C}}$ is opposite. 
The calculated structure factors are listed in Table S-I. 
The relatively poorer agreement at the $E1$ resonance energy of 6.720 keV is due to the imperfect correction of absorption at the edge. 

\begin{table}[b]
\caption{Crystal structure factor $F_{\text{C}}$ in the field-induced phase at 15 K and 6 T for $H\parallel [1 \bar{1} 0]$ calculated from the atomic displacement parameters reported in Ref. 22. }
\begin{tabular}{ccc}
\hline
$h$ $k$ $l$ & & $F_{\text{C}}$ \\
\hline
1 1 1 & &  $0.204 - 0.00936 i$ \\
3 3 3 & &  $-0.331 + 0.0232 i$ \\
2 2 3 & &  $0.608 - 0.0866 i$ \\
2 2 1 & &  $-0.115 + 0.0227 i$ \\
2 2 5 & &  $-0.912 + 0.148 i$ \\
\hline
\end{tabular}
\end{table}

\newpage
\section*{XMCD at the Samarium $\bm{L_2}$-edge}
\begin{figure}[b]
\begin{center}
\includegraphics[width=8cm]{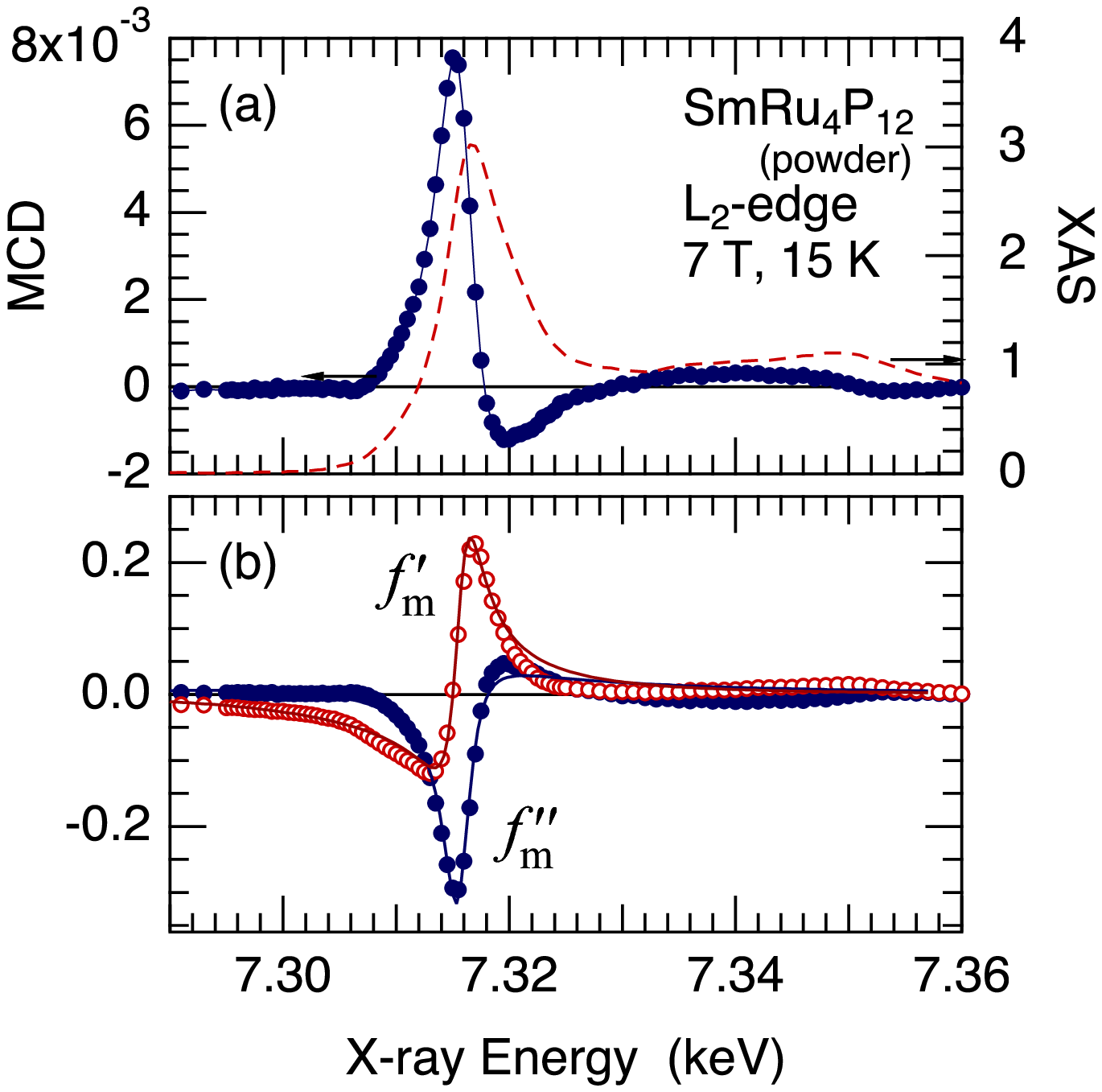}
\end{center}
\caption{(a) XMCD (filled circles) and XAS (dashed line) of SmRu$_4$P$_{12}$ at 15 K and 7 T in phase II around the Sm $L_2$-edge.  (b) Real and imaginary parts of the magnetic dipolar spectral function $f_{\text{m}}(\omega)$ in an arbitrary unit. 
Solid lines are the fits using Eq.~(3) in the main text.}
\label{fig:MCDL2}
\end{figure}

Figure~\ref{fig:MCDL2}(a) shows the XMCD and XAS spectra around the Sm $L_2$-edge at 15 K and 7 T in phase II. 
In contrast to the result for the $L_3$-edge, the XMCD spectrum exhibits a large anomaly at the $E1$-transition energy whereas there is little anomaly at the $E2$-transition energy. 
This could be understood by the dominant contribution of the $E1$ resonance to the XMCD of Sm compounds at the $L_2$ edge, which mainly reflects the Sm $4f$--$5d$ exchange interaction as explained in Ref.~40. 
As described in the main text, we directly deduced the imaginary part $f''_{\text{m}}(\omega)$ from the XMCD and XAS spectra. 
The real part $f'_{\text{m}}(\omega)$ was obtained by the Kramers-Kronig transformation. 
$f'_{\text{m}}(\omega)$ and $f''_{\text{m}}(\omega)$ thus obtained are shown in Fig.~\ref{fig:MCDL2}(b). 

\newpage
\section*{RSXD at the Samarium $\bm{M_{4,5}}$-edge}
\begin{figure}[b]
\begin{center}
\includegraphics[width=7.5cm]{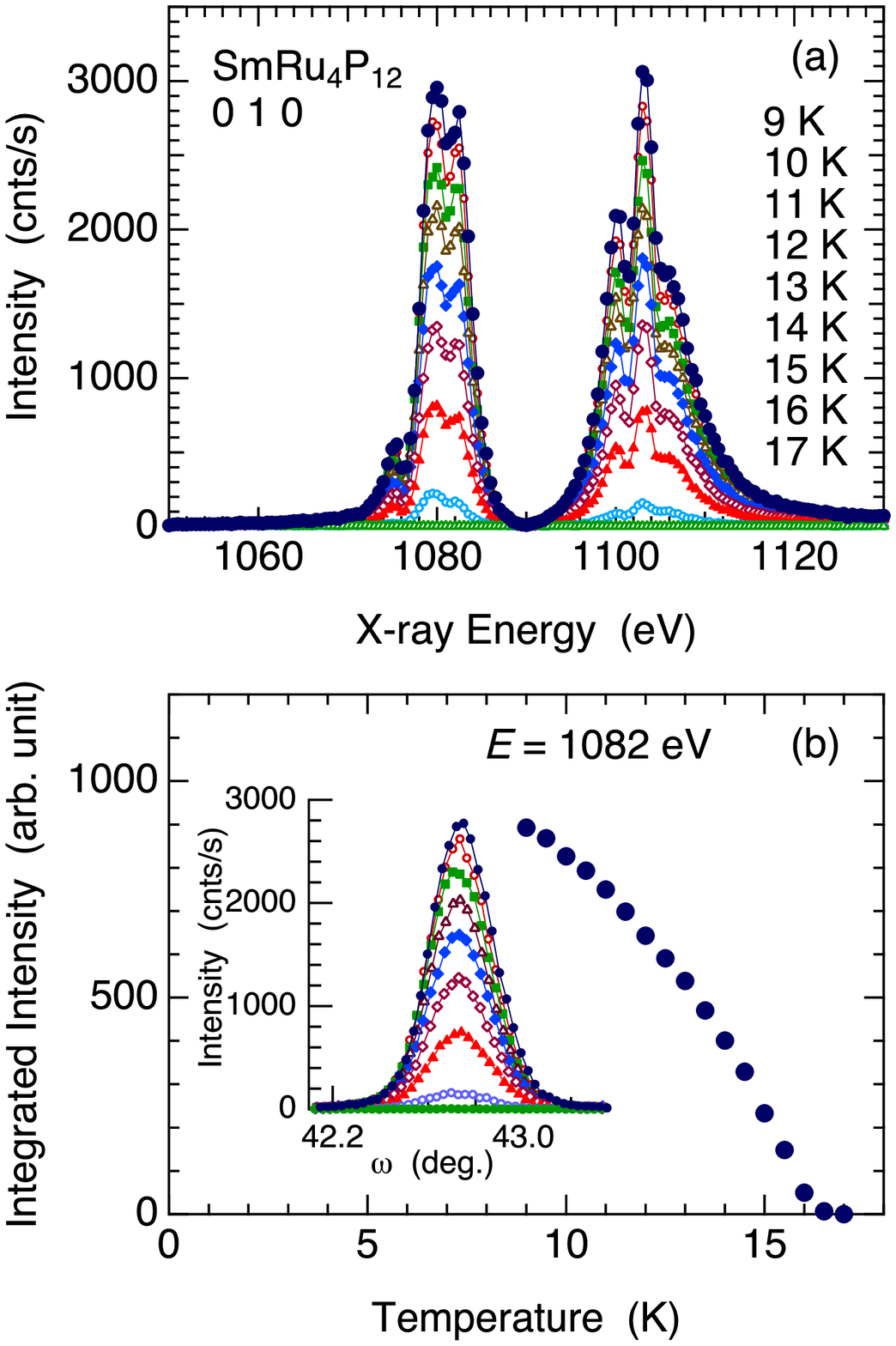}
\end{center}
\caption{Results of resonant soft x-ray diffraction at the Sm $M$-edge for a $\pi$-polarized incident beam without analyzing the final polarization. 
(a) X-ray energy dependences of the 010 Bragg intensity at zero field from 9 K to 17 K. 
(b) Temperature dependence of the integrated intensity of the resonant Bragg diffraction at 1082 eV ($M_5$-edge). 
Inset shows the rocking scans of the Bragg peak from 9 K to 17 K. 
}
\label{fig:RSXD_SmMedge}
\end{figure}

Figure~\ref{fig:RSXD_SmMedge}(a) shows the energy dependences of the 010 Bragg intensity at the Sm $M$-edge. 
The measurement was performed at zero field without polarization analysis. 
The intensity exhibits resonant enhancements around 1080 eV ($M_5$-edge) and around 1104 eV ($M_4$-edge). 
Both resonances consist of some fine structures. 
Since the $M$-edge resonance arises from the $3d$--$4f$ dipole transition, the resonant signal directly reflects the $4f$ electronic state. 
At zero field, where only the AFM order exists, the resonance is purely of magnetic origin. 
Figure~\ref{fig:RSXD_SmMedge}(b) shows the temperature dependence of the integrated intensity for the rocking scans shown in the inset. 
The temperature dependence, which directly reflects the AFM order of the $4f$ state, is exactly the same as those obtained at the Sm $L$-edge and the P $K$-edge shown in Fig.~4 and Fig.~9 in the main text, respectively.  

\newpage


\begin{thebibliography}{99}
\bibitem{comment1} To be exact, in the region of local expansion, the nearest neighbor P--P distance in the P$_{12}$ icosahedron increases, keeping the distance from the Sm atom at the center almost unchanged, resulting in a slight increase in the volume of the P$_{12}$ icosahedron. 
The relative expansion of the Ru cube, in contrast, is 10 times larger than that of the P$_{12}$ icosahedron because the Ru atoms shift along the radial direction from the Sm atom at the center.
\end{thebibliography}
\end{document}